\pdfoutput=1
\documentclass[journal]{IEEEtran}
\PassOptionsToPackage{table}{xcolor}
\usepackage{cite}
\usepackage{amsmath,amssymb}
\usepackage{graphicx,color}
\usepackage{textcomp}
\usepackage{array}
\usepackage{booktabs}
\AtBeginDocument{\definecolor{acrzeb}{rgb}{0.92,0.95,0.99}}
\usepackage{tikz}
\usetikzlibrary{arrows.meta,calc,shadows}
\AtBeginDocument{\definecolor{headblue}{cmyk}{0.93,0.59,0.15,0.02}}
\newcommand{\eff}{\mathrm{eff}}
\newcommand{\SINR}{\gamma}
\usepackage{pifont}
\usepackage{multirow}
\DeclareRobustCommand{\cmark}{\ding{51}}
\DeclareRobustCommand{\xmark}{\ding{55}}
\DeclareRobustCommand{\pmark}{\tikz[baseline=-0.55ex]{\draw[line width=0.5pt] (0,0) circle (0.75ex);\fill (0,0.75ex) arc (90:270:0.75ex) -- cycle;}}
\usepackage{amsthm}
\newtheoremstyle{remarkit}{\topsep}{\topsep}{\itshape}{}{\itshape}{.}{5pt plus 1pt minus 1pt}{}
\theoremstyle{remarkit}
\newtheorem{remark}{Remark}

\begin{document}

\title{Multiple Access in Pinching-Antenna Systems: A Survey, Impairment Analysis, and Evaluation}

\author{Wei Jiang, \IEEEmembership{Senior Member, IEEE}, and Hans Dieter Schotten
\thanks{The authors are with the Intelligent Networking Research Group, German Research Center for Artificial Intelligence (DFKI), 67663 Kaiserslautern, Germany, and also with the Department of Electrical and Computer Engineering, University of Kaiserslautern (RPTU), 67663 Kaiserslautern, Germany, e-mail: (wei.jiang@dfki.de; schotten@eit.uni-kl.de).}
\thanks{References cited as ``early access'' or by arXiv identifier were verified against the publisher's records as of submission; their final bibliographic details will be updated at the proof stage.}}

\maketitle

\begin{abstract}
Pinching-antenna systems attach dielectric pinches to a waveguide, creating line-of-sight radiation points within a few meters of any user. The pinches on one waveguide, however, radiate the same signal, unlike a conventional array, where each element transmits a dedicated signal. Separating the signals of multiple users therefore becomes challenging and requires a new perspective on multiple access. Most existing studies rest on an idealized model, and no survey examines the access schemes under realistic impairments. This article surveys multiple access in pinching-antenna systems and organizes fifty-six works into a unified taxonomy by spatial degrees of freedom. The users of one waveguide can be separated only in time, frequency, power, or guided mode, while multiple waveguides add the space domain. The article then identifies the impairments critical to this separation, which perturb the effective channel, leak interference across orthogonal resources, or couple the pinches on the uplink, and summarizes them in an impairment-versus-method map. A unified evaluation of time-, frequency-, power-, and space-domain access under identical channel and impairment models reveals that the degradation is method specific. The analytical and numerical results give insight into multiple access in these systems and provide a basis for its impairment-aware design, a necessary step toward their adoption in practical networks.
\end{abstract}


\begin{table*}[!t]
\centering
\caption{List of Acronyms}
\label{tab:acronyms}
\footnotesize
\renewcommand{\arraystretch}{1.05}
\setlength{\arrayrulewidth}{0.4pt}
\rowcolors{2}{acrzeb}{white}%
\begin{tabular}{@{}>{\bfseries}l@{\hskip 8pt}p{6.7cm}@{\hskip 16pt}>{\bfseries}l@{\hskip 8pt}p{6.7cm}@{}}
\toprule
\rowcolor{headblue}\textcolor{white}{Acronym} & \textcolor{white}{\textbf{Definition}} & \textcolor{white}{Acronym} & \textcolor{white}{\textbf{Definition}} \\
\midrule
ADC    & analog-to-digital converter                   & OMA    & orthogonal multiple access \\
BER    & bit-error rate                                & PA     & pinching antenna \\
CDMA   & code-division multiple access                 & PASS   & pinching-antenna system \\
CoMP   & coordinated multipoint                        & RF     & radio frequency \\
CSI    & channel-state information                     & RIS    & reconfigurable intelligent surface \\
DAC    & digital-to-analog converter                   & RSMA   & rate-splitting multiple access \\
DL     & downlink                                      & SA     & segment aggregation \\
DoF    & degrees of freedom                            & SC-FDE & single-carrier frequency-domain equalization \\
EDMA   & environment-division multiple access          & SCMA   & sparse code multiple access \\
FDMA   & frequency-division multiple access            & SDMA   & space-division multiple access \\
IAR    & inter-antenna radiation                       & SIC    & successive interference cancellation \\
IDMA   & interleave-division multiple access           & SINR   & signal-to-interference-plus-noise ratio \\
ISI    & inter-symbol interference                     & SM     & segment multiplexing \\
IWA    & in-waveguide attenuation                      & SNR    & signal-to-noise ratio \\
LO     & local oscillator                              & SS     & segment selection \\
LoS    & line-of-sight                                 & SWAN   & segmented waveguide-enabled pinching-antenna system \\
MA     & multiple access                               & TDMA   & time-division multiple access \\
MDM    & mode-division multiplexing                    & UL     & uplink \\
MIMO   & multiple-input multiple-output                & URLLC  & ultra-reliable low-latency communications \\
MISO   & multiple-input single-output                  & WD     & waveguide division \\
NLoS   & non-line-of-sight                             & WDMA   & waveguide-division multiple access \\
NOMA   & non-orthogonal multiple access                & WG     & waveguide \\
OFDM   & orthogonal frequency-division multiplexing    & WM     & waveguide multiplexing \\
OFDMA  & orthogonal frequency-division multiple access & WS     & waveguide switching \\
\end{tabular}
\end{table*}

\section{Introduction}
\IEEEPARstart{F}{lexible}-antenna technologies---reconfigurable intelligent surfaces (RISs)~\cite{Ref_Wu2020RIS}, fluid antennas~\cite{Ref_Wong2021FluidAntenna}, movable antennas~\cite{Ref_Zhu2024MovableAntenna}, and dynamic metasurface antennas (DMAs)~\cite{Ref_Shlezinger2021DMA}---reshape the wireless channel but cannot defeat large-scale path loss, because their radiating elements move only on the order of a wavelength (or, for DMAs, not at all, within a fixed aperture) and, in the RIS case, suffer a cascaded (multiplicative) path loss. Pinching-antenna systems (PASS), prototyped first by NTT DOCOMO \cite{Ref_Fukuda2022Pinching} and analyzed for communications by Ding \emph{et al.}\ \cite{Ref_Ding2025PinchingPerspective}, remove this limitation. A long dielectric waveguide carries the signal for tens of meters with only a few decibels of in-waveguide attenuation (IWA) at the millimeter-wave bands of interest \cite{Ref_Xu2025InWGAtten}; small dielectric ``pinches'' (pinching antennas, PAs) clipped onto the guide each leak a controllable fraction of the guided wave to free space at a \emph{chosen} location. A PA can therefore be instantiated within a few meters of any user, converting the access link into a strong, reconfigurable line-of-sight (LoS) near-field channel and turning large-scale path loss from a fixed obstacle into a \emph{programmable} parameter.

What sets PASS apart is the \emph{reach} of its reconfigurability: a radiating element can be placed \emph{anywhere} along a waveguide that spans tens of meters, so PASS not only shortens the access distance but also routes around non-line-of-sight (NLoS) blockage---moving the coverage problem from the ``last mile'' to the ``last meter.'' The fundamentals were established in single-user settings---a simplification that produced the field's seminal results \cite{Ref_Ding2025PinchingPerspective,Ref_Ouyang2025ArrayGain,Ref_Tyrovolas2026Performance}---but a practical system must serve many users at once, and there the same meter-scale placement range makes multiple-access (MA) behavior fundamentally different from a co-located array's: each element of such an array has its own radio-frequency (RF) chain and precoded signal, whereas the PAs on a waveguide are spread over meters yet share a single RF chain and radiate the same composite signal.

\begin{figure*}[!t]
\centering
\resizebox{\textwidth}{1.2\height}{%
\begin{tikzpicture}[>=Stealth, font=\sffamily\scriptsize,
  chip/.style={draw, rounded corners=2pt, inner xsep=4.5pt, inner ysep=2.4pt,
               font=\sffamily\bfseries\tiny, text=black!85, line width=0.4pt,
               top color=white, bottom color=black!10,
               drop shadow={shadow xshift=0.5pt, shadow yshift=-0.5pt, fill=black, opacity=0.22}},
  chipb/.style={chip, draw=blue!45!black, bottom color=blue!16},
  chipo/.style={chip, draw=orange!55!black, bottom color=orange!22},
  chipt/.style={chip, draw=teal!45!black, bottom color=teal!15},
  chipg/.style={chip, draw=black!45, text=black!55, bottom color=black!8}]
  \draw[fill=blue!4, draw=blue!55!black, line width=0.6pt, rounded corners=3pt,
        drop shadow={shadow xshift=1pt, shadow yshift=-1pt, fill=black, opacity=0.15}]
        (0.3,0.55) rectangle (6.5,2.28);
  \begin{scope}
    \clip[rounded corners=3pt] (0.3,0.55) rectangle (6.5,2.28);
    \fill[blue!55!black] (0.3,1.86) rectangle (6.5,2.28);
    \fill[blue!12] (0.3,1.54) rectangle (6.5,1.86);
  \end{scope}
  \node[white, font=\sffamily\bfseries\footnotesize] at (3.4,2.07) {Single-Waveguide};
  \node[text=blue!40!black, font=\sffamily\itshape\tiny] at (3.4,1.70) {$M{=}1$, one composite signal per RF chain};
  \node[chipb] at (1.30,1.30) {TDMA};
  \node[chipb] at (3.05,1.30) {FDMA / OFDMA};
  \node[chipb] at (4.85,1.30) {NOMA};
  \node[chipb] at (2.30,0.82) {MDM};
  \node[chipb] at (3.85,0.82) {SWAN-SS/SA};
  \draw[left color=blue!25, right color=orange!55, draw=black!35, line width=0.4pt, line join=round,
        drop shadow={shadow xshift=0.6pt, shadow yshift=-0.6pt, fill=black, opacity=0.15}]
        (6.62,1.26) .. controls (7.80,1.29) .. (8.95,1.38)
        -- (8.95,1.52) -- (9.48,1.20) -- (8.95,0.88) -- (8.95,1.02)
        .. controls (7.80,1.11) .. (6.62,1.14) -- cycle;
  \node[align=center, font=\sffamily\itshape\tiny, text=orange!50!black] at (8.05,1.80) {increasing spatial\\degrees of freedom};
  \draw[fill=orange!6, draw=orange!55!black, line width=0.6pt, rounded corners=3pt,
        drop shadow={shadow xshift=1pt, shadow yshift=-1pt, fill=black, opacity=0.15}]
        (9.5,0.55) rectangle (15.7,2.28);
  \begin{scope}
    \clip[rounded corners=3pt] (9.5,0.55) rectangle (15.7,2.28);
    \fill[orange!55!black] (9.5,1.86) rectangle (15.7,2.28);
    \fill[orange!14] (9.5,1.54) rectangle (15.7,1.86);
  \end{scope}
  \node[white, font=\sffamily\bfseries\footnotesize] at (12.6,2.07) {Multi-Waveguide};
  \node[text=orange!45!black, font=\sffamily\itshape\tiny] at (12.6,1.70) {$M\ge 2$, multiple RF chains: adds spatial DoF};
  \node[chipo] at (10.45,1.30) {SDMA / WM};
  \node[chipo] at (12.05,1.30) {WDMA / WD};
  \node[chipo] at (13.45,1.30) {EDMA};
  \node[chipo] at (14.95,1.30) {SWAN-SM};
  \node[chipo] at (11.85,0.82) {CoMP};
  \node[chipo] at (13.55,0.82) {cell-free};
  \draw[fill=teal!5, draw=teal!45!black, line width=0.6pt, rounded corners=3pt,
        drop shadow={shadow xshift=1pt, shadow yshift=-1pt, fill=black, opacity=0.15}]
        (4.4,-1.62) rectangle (11.6,-0.50);
  \begin{scope}
    \clip[rounded corners=3pt] (4.4,-1.62) rectangle (11.6,-0.50);
    \fill[teal!45!black] (4.4,-0.92) rectangle (11.6,-0.50);
  \end{scope}
  \node[white, font=\sffamily\bfseries\footnotesize] at (8.0,-0.71) {Emerging / Cross-Regime};
  \node[chipt] at (6.65,-1.27) {RSMA (single- and multi-WG)};
  \node[chipt] at (9.95,-1.27) {index modulation};
  \draw[teal!45!black, dashed, line width=0.6pt, ->] (5.6,-0.48) to[bend left=12] (5.2,0.51);
  \draw[teal!45!black, dashed, line width=0.6pt, ->] (10.4,-0.48) to[bend right=12] (10.8,0.51);
  \draw[fill=black!2, draw=black!50, dashed, line width=0.5pt, rounded corners=3pt]
        (12.2,-1.62) rectangle (15.7,-0.50);
  \node[text=black!60, font=\sffamily\bfseries\footnotesize] at (13.95,-0.73) {Absent to date};
  \node[chipg] at (13.95,-1.27) {code-domain};
\end{tikzpicture}}
\caption{Taxonomy of PASS multiple access, organized along the spatial-DoF spine: the waveguide-to-user ratio sets whether spatial-domain access is possible, and the single-waveguide case is the hardest-constrained building block. The emerging families span both regimes (dashed arrows).}
\label{fig:tax}
\end{figure*}
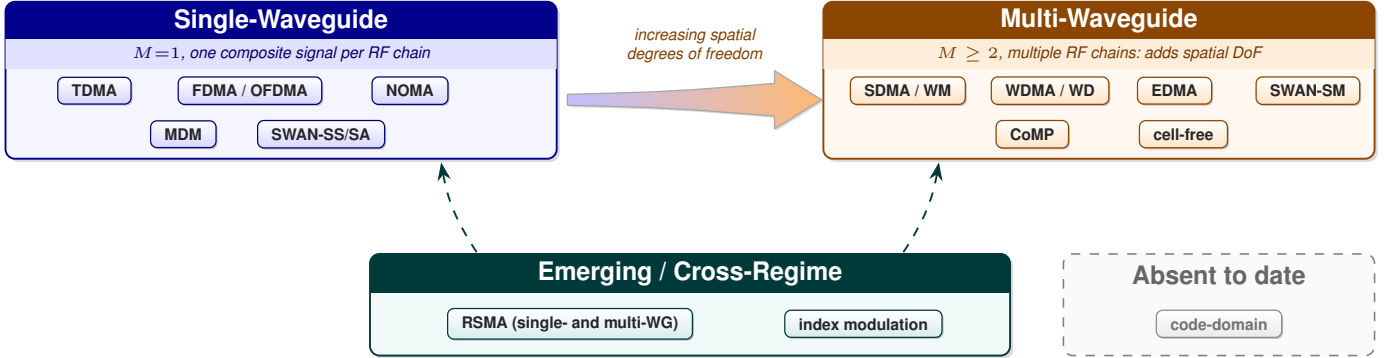

Multiple access is therefore not just one PASS feature among many: it decides whether pinching's last-meter gain---clear for a single user---carries over once many users share one waveguide. Two properties shape the answer---the defining constraint of PASS and its defining freedom---and each reshapes the MA mechanism in its own way:
\begin{itemize}
\item \emph{The constraint: one waveguide carries one composite signal.} Every PA re-radiates a phase-shifted, attenuated replica of the same feed signal, so there is no per-user precoder and \emph{no spatial-domain multiple access} on a single guide---users must be separated in time, frequency, power, or guided mode. The shared feed further imposes structural penalties a co-located array never pays: the transmit power is diluted across the active PAs; one phase profile must serve every user at once, a multi-user phase-alignment conflict; and, on the uplink, the wave captured by one PA is partly re-radiated by the others (\emph{inter-antenna radiation}, IAR~\cite{Ref_Gu2025SWANUplink}).
\item \emph{The freedom: pinching beamforming couples geometry to resource allocation.} Moving or activating PAs reshapes every user's path loss and phase at once, so the access scheme and the antenna positions cannot be optimized separately. And because repositioning is \emph{mechanical} rather than electronic, every reconfiguration carries a latency far longer than any electronically steered array encounters, together with timing jitter---a cost that falls hardest on time-division access, the scheme that re-pinches most often.
\end{itemize}
The costs of both properties are one reason why the central debate between non-orthogonal multiple access (NOMA) and orthogonal multiple access (OMA)~\cite{Ref_Ren2025NOMAorOMA} receives opposite answers: each study idealizes them in its own way---a gap Section~\ref{ssec:gap} makes precise.

\subsection{State of the Art}\label{sec:catalog}
Multiple access has rapidly become one of the most active threads of PASS research: the fifty-six works of Table~\ref{tab:cmp} now touch on it, though many adopt a multi-user setting rather than making multiple access their central focus. To bring order to this fast-growing yet fragmented literature, we anchor this article to a single structural fact: \emph{the spatial degrees of freedom (DoF) of PASS multiple access are set by the waveguide-to-user ratio} (Fig.~\ref{fig:tax}). With one RF-fed waveguide, spatial separation is impossible and the system is forced into time-, frequency-, power-, or mode-domain access; the single waveguide is the atomic, hardest-constrained building block. With multiple waveguides---each its own feed---space-division multiple access (SDMA) and its hybrids become available, at the price of inter-waveguide coupling and multi-chain synchronization. Fig.~\ref{fig:tax} therefore maps the related work along this spine: the single-waveguide regime; the multi-waveguide regime; and a set of cross-regime and emerging families. Table~\ref{tab:cmp} summarizes the surveyed works family by family, together with the impairments each models. We catalog each family at the level of its core mechanism and defining results, deferring its behavior under realistic impairments---among them re-pinching latency, beam squint, position error, and imperfect successive interference cancellation (SIC)---to the impairment-aware core (Sections~\ref{sec:impair}--\ref{sec:bench}, where the impairments are in turn identified, analyzed, and benchmarked). The structural penalties of the shared feed---per-antenna power dilution and the multi-user phase-alignment conflict on the downlink, and inter-antenna radiation on the uplink (Section~\ref{sec:model})---are carried throughout.

\definecolor{famcol}{rgb}{0.93,0.93,0.93}
\newcommand{\cmpgrp}[3][0pt]{\multirow{-#2}{*}{\raisebox{#1}[0pt][0pt]{\rotatebox[origin=c]{90}{\textbf{\textcolor{white}{#3}}}}}}
\newcommand{\cmphl}{\cellcolor{acrzeb}}
\newcommand{\fc}{\cellcolor{famcol}}
\newcommand{\cmphd}[1]{{\footnotesize\bfseries\color{white}#1}}%
\newcommand{\famsep}{\arrayrulecolor{headblue!70}\cline{1-1}\arrayrulecolor{black!35}\cline{2-7}\arrayrulecolor{black}}
\begin{table*}[!p]
\renewcommand{\arraystretch}{0.90}
\setlength{\tabcolsep}{4pt}
\setlength{\aboverulesep}{0.8pt}\setlength{\belowrulesep}{0.8pt}
\DeclareRobustCommand{\xmark}{\scalebox{1.2}{\ding{55}}}%
\caption{Comparison of PASS Multiple-Access Works by Family. The \emph{Impairments modeled} column gives which impairments each work models; \xmark{} = none (idealized, or only generalized impairments outside our scope); ``--'' = unstated.}
\label{tab:cmp}
\centering
\scriptsize
\begin{tabular}{>{\columncolor{headblue!70}}c l c c c c p{8.8cm}}
\toprule
\rowcolor{headblue}\rule[-5pt]{0pt}{16pt} & \cmphd{Family} & \cmphd{Ref} & \cmphd{Link} & \cmphd{Users} & \cmphd{Impairments modeled} & \multicolumn{1}{l}{\cmphd{Core idea / key result}} \\
\midrule
 & \fc & \cmphl\cite{Ref_Ren2025NOMAorOMA} & \cmphl DL & \cmphl $2$ & \cmphl \xmark & \cmphl TDMA outperforms NOMA for symmetric rate targets; NOMA outperforms FDMA \\
 & \fc & \cite{Ref_Chen2025SumCapacityMAC} & UL & $K$ & \xmark & Alternating transmission per user achieves the MAC sum capacity; SIC unnecessary \\
 & \fc & \cmphl\cite{Ref_Tegos2025UplinkMinRate} & \cmphl UL & \cmphl $K$ & \cmphl \xmark & \cmphl Max-min fair via closed-form Lambert-$W$ allocation; positions fixed per frame \\
 & \fc & \cite{Ref_Liu2025WPTMECTDMAvsNOMA} & UL & $3$ & discrete activation & TDMA is superior to NOMA for wireless-powered MEC under dynamic activation \\
 & \fc & \cmphl\cite{Ref_Xie2025LowComplexityPlacement} & \cmphl DL & \cmphl $K$ & \cmphl \xmark & \cmphl Closed-form placements; single-PA TDMA outperforms NOMA, which hurts edge users \\
 & \fc\multirow{-6}{*}{TDMA} & \cite{Ref_Xie2025BGATGNN} & DL & $2$--$5$ & \xmark & Graph-attention placement--power design; fast, size-generalizable inference \\
\famsep
 & \raisebox{-4pt}[0pt][0pt]{OFDMA} & \cmphl\cite{Ref_Oikonomou2025OFDMA} & \cmphl DL & \cmphl $2$, $4$ & \cmphl \xmark & \cmphl PAs meters apart form a multi-tap channel whose ISI single-carrier designs overlook; max-min OFDMA (greedy subcarriers $+$ water-filling) beats TDMA/SC-FDE \\
\famsep
 & \fc & \cite{Ref_Ding2025PinchingPerspective} & DL & $2$ & \xmark & Founding single-WG NOMA; placement sets the SIC gain ordering \\
 & \fc & \cmphl\cite{Ref_Ding2025AnalyticalPlacement} & \cmphl DL & \cmphl $K$; $2$ & \cmphl \xmark & \cmphl Analytical placement optimization for OMA (max-min/greedy) and NOMA (power-min) \\
 & \fc & \cite{Ref_Wang2025AntennaActivationNOMA} & DL & $2$, $20$ & IWA & Discrete PA activation via matching; $\approx$99\% of exhaustive sum rate \\
 & \fc & \cmphl\cite{Ref_Zhang2026EEDiscreteActivation} & \cmphl DL & \cmphl $2$, $4$, $5$ & \cmphl activation power & \cmphl EE maximization for discrete-PA NOMA under per-PA activation power \\
 & \fc & \cite{Ref_Mohammadzadeh2025PowerAllocNOMA} & DL & $4$, $8$ & \xmark & Total transmit-power min.\ under per-user QoS; recursive closed-form allocation \\
 & \fc & \cmphl\cite{Ref_Mohammadzadeh2025CNNPower} & \cmphl DL & \cmphl $2$--$10$ & \cmphl \xmark & \cmphl CNN power-allocation surrogate; inference complexity independent of $K$ \\
 & \fc & \cite{Ref_Vashakidze2025RadiationPower} & DL & $3$--$8$ & SIC order & Sum-rate max.\ over radiated power (fixed PAs) and user power allocation; $+$10--15\% vs.\ equal power; dynamic SIC reordering near-optimal \\
 & \fc & \cmphl\cite{Ref_Zhou2025SumRateNOMA} & \cmphl DL & \cmphl $2$ & \cmphl \xmark & \cmphl Sum-rate max.\ jointly over power (closed-form KKT) and positions (bisection) \\
 & \fc & \cite{Ref_Xu2025QoSCRNOMA} & DL & $2$ & \xmark & CR-NOMA with primary-QoS guarantee; $N{=}1$ solved globally in closed form \\
 & \fc & \cmphl\cite{Ref_Cheng2025OMANOMAPerformance} & \cmphl DL & \cmphl $2$ & \cmphl \xmark & \cmphl Closed-form outage in mixed LoS/NLoS rooms; NLoS user keeps diversity order one \\
 & \fc & \cite{Ref_Cheng2026SingleWGOMAvsNOMA} & UL+DL & $K$ & \xmark & Closed-form outage probability and ergodic rate, OMA vs.\ NOMA (UL/DL); OMA generally superior at high SNR \\
 & \fc & \cmphl\cite{Ref_Yue2026NOMAPerformance} & \cmphl DL & \cmphl $2$ & \cmphl imperfect SIC & \cmphl Stochastic-geometry blockage probability and ergodic rate over LoS/NLoS links \\
 & \fc & \cite{Ref_AlaaEldin2026BERNOMA} & UL+DL & $2$ & SIC, position error & Exact BER under imperfect SIC; PA position optimized to minimize BER \\
 & \fc & \cmphl\cite{Ref_Zeng2025ULSumRateNOMA} & \cmphl UL & \cmphl $3$ & \cmphl \xmark & \cmphl Sum-rate max.\ via joint position--power optimization, $\approx$ exhaustive optimum; NOMA outperforms TDMA (single PA) \\
 & \fc & \cite{Ref_Zeng2025ULEENOMA} & UL & $K$ & \xmark & EE maximization via joint position--power optimization; the optimum serves only the strongest user(s), needing a QoS guard \\
 & \fc & \cmphl\cite{Ref_Ye2026FairnessNOMACSI} & \cmphl DL & \cmphl $K$ & \cmphl imperfect SIC & \cmphl Robust max-min SINR via joint PA position and NOMA power; conservative SINR keeps SIC feasible under imperfect CSI; $+$7--10\,dB vs.\ non-robust \\
 & \fc & \cite{Ref_Chi2025NOMAEavesdropping} & DL & $2$ & \xmark & Near-user eavesdropping; coupling-length control zeroes the secrecy outage \\
 & \fc & \cmphl\cite{Ref_Ai2026NOMAMECDelay} & \cmphl UL & \cmphl $2$ & \cmphl SIC order & \cmphl Min-max offloading delay; repositioning reshuffles SIC order, forcing $O(K!)$ search \\
 & \fc & \cite{Ref_Pakravan2026AirCompNOMA} & UL & $4{+}3$ & \xmark & Over-the-air computation coexists with NOMA; its signal is non-cancelable by SIC \\
 & \fc & \cmphl\cite{Ref_Ahmed2026SemanticNOMA} & \cmphl DL & \cmphl $2$ & \cmphl \xmark & \cmphl Semantic$+$bit pair under bit-to-semantic SIC; 6--15\% semantic-SE gain \\
 & \fc\multirow{-21}{*}{NOMA} & \cite{Ref_Zhuo2026PNOMA} & DL & $2$ & \xmark & Overlap ratios bridge OMA$\leftrightarrow$NOMA and shift the PA optimum; highest sum rate \\
\famsep
 & \multirow{2}{*}[-3.5pt]{MDM$^\ast$} & \cmphl\cite{Ref_Xu2026MultiModeMDM} & \cmphl DL & \cmphl $2$ & \cmphl mode crosstalk & \cmphl Independent streams via orthogonal guided modes (one RF chain each); zero-forcing absorbs mode leakage; outperforms single-mode TDMA \\
\famsep
 & \fc & \cite{Ref_Ouyang2025SWAN} & UL+DL & $1$ & IAR, IWA & Foundational segmented-waveguide framework; segment selection (SS), aggregation (SA), and multiplexing (SM) feeds; SM$>$SA$>$SS \\
\cmpgrp[36.7pt]{31}{Single-waveguide} & \fc\multirow{-2}{*}{SWAN} & \cmphl\cite{Ref_Gu2025SWANUplink} & \cmphl UL & \cmphl $4$ & \cmphl IAR, IWA & \cmphl Segmented guide removes uplink inter-antenna radiation \\
\midrule
 & \fc & \cite{Ref_Bereyhi2025MIMO} & DL & $K$ & \xmark & Joint optimization of the precoder and PA locations maximizes weighted sum rate \\
 & \fc & \cmphl\cite{Ref_MIMOPASS2025DLUL} & \cmphl UL+DL & \cmphl $K$ & \cmphl \xmark & \cmphl Joint beamforming extends to UL, multi-PA guides (IAR-neglecting); UL exceeds DL \\
 & \fc & \cite{Ref_Zhao2025TransmissionStructures} & DL & $K$ & \xmark & Waveguide multiplexing/division/switching (WM/WD/WS) structures; max-min-fair design; WS suits unicast, WM multicast; WM--WD gap narrows for separated users \\
 & \fc\multirow{-4}{*}[4pt]{SDMA} & \cmphl\cite{Ref_Zhang2025ULMISO} & \cmphl UL & \cmphl $K$ & \cmphl \xmark & \cmphl Uplink multi-user MISO across guides; joint PA placement and user power \\
\famsep
 & \multirow{3}{*}[-13pt]{WDMA} & \cite{Ref_Ding2025WDMA} & DL & $K$ & \xmark & One waveguide per user as a new radio resource; joint power and pinching-beamforming design achieves near-orthogonality in free space \\
 & & \cmphl\cite{Ref_Zhang2026WDMAvsNOMA} & \cmphl DL & \cmphl $2$ & \cmphl \xmark & \cmphl Idealized closed-form comparison for downlink outage probability/average rate: NOMA leads at high SNR, WDMA at low but hits an interference floor \\
 & & \cite{Ref_Wang2025MultiWGActNOMA} & DL & $K$ & \xmark & Multi-WG NOMA at discrete PA positions; joint guide assignment, activation, decoding order, and power; SIC order $\propto(I{+}\sigma^2)/|g|^2$, power-allocation-invariant \\
\famsep
 & \fc\raisebox{-5pt}[0pt][0pt]{EDMA} & \cmphl\cite{Ref_Ding2025EDMA} & \cmphl UL+DL & \cmphl $K$ & \cmphl \xmark & \cmphl Blockage reconfiguration as a new MA domain; closed-form UL/DL gain and win-probability over PA-assisted TDMA ($K{=}2$); near-optimal PA placement \\
\famsep
 & & \cite{Ref_Amhaz2025CoMP} & DL & $K$ & \xmark & Two-BS joint-transmission CoMP; joint beamforming and PA placement under QoS \\
 & & \cmphl\cite{Ref_Li2025CellFree} & \cmphl DL & \cmphl $K$ & \cmphl \xmark & \cmphl Cell-free PASS outperforms cell-free MIMO; rate loss curbed as users grow \\
 & & \cite{Ref_Sun2025MultiCellSG} & DL & -- & \xmark & Stochastic-geometry outage probability analysis under blockage; interference statistics match fixed antennas, so pinching does not raise the interference floor \\
 & & \cmphl\cite{Ref_Zhu2026Topology} & \cmphl DL & \cmphl $K$ & \cmphl \xmark & \cmphl Voronoi-topology analysis of large-scale deployment; closed-form 1-D/2-D excitation positions; cell layout becomes more uniform, lifting average rates \\
\cmpgrp[27pt]{13}{Multi-waveguide} & \multirow{-5}{*}{Multi-cell} & \cite{Ref_Ding2026QuietWorld} & DL & $K$ & \xmark & Multi-cell power minimization via cross-entropy PA position optimization: QoS is met by ``whispering'' rather than ``shouting'', without inter-BS data sharing or synchronization \\
\midrule
 & \fc & \cmphl\cite{Ref_RSMA2025MultiWG} & \cmphl DL & \cmphl $K$ & \cmphl \xmark & \cmphl Sum-rate maximization; RSMA $>$ NOMA across waveguides \\
 & \fc\multirow{-2}{*}{RSMA} & \cite{Ref_RSMA2025Uplink} & UL & $2$ & \xmark & Closed-form outage probability analysis; rate splitting removes the NOMA outage floor \\
\famsep
 & \multirow{2}{*}{Index mod.$^\dagger$} & \cmphl\cite{Ref_Yang2025PAIM} & \cmphl p2p & \cmphl -- & \cmphl \xmark & \cmphl Bits conveyed by PA-position index (point-to-point) \\
\cmpgrp{4}{Emerging} & & \cite{Ref_PASM2026SpatialMux} & p2p & -- & \xmark & PA-position phases give a phase-shifter-free composite constellation \\
\midrule
 & & \cmphl\cite{Ref_Xu2025InWGAtten} & \cmphl DL & \cmphl $1$; $K$ & \cmphl IWA & \cmphl Rate loss grows as the square of IWA, which depends on waveguide material, structure, and frequency; negligible at $0.08$\,dB/m \\
 & & \cite{Ref_Ouyang2025ArrayGain} & DL & $1$ & coupling & Array gain non-monotone in antenna number and inter-PA spacing \\
 & & \cmphl\cite{Ref_Tyrovolas2026Performance} & \cmphl DL & \cmphl $1$ & \cmphl IWA & \cmphl Closed-form outage probability and average rate; IWA-aware optimal placement \\
 & & \cite{Ref_Wang2025Modeling} & DL & $K$ & coupling & Physics-based hardware model via coupled-mode theory; equal- and proportional-power models proposed, near-identical; power cut $>$95\% vs.\ massive MIMO \\
 & & \cmphl\cite{Ref_Zubair2026Pattern} & \cmphl -- & \cmphl -- & \cmphl directivity & \cmphl Closed-form directional pattern; omni placements lose ${\sim}7.5$\,dB \\
 & & \cite{Ref_Zhang2025DiPASS} & DL & $K$ & directivity, IWA & Channel model with directional radiation and $1.3$\,dB/m in-waveguide loss; closed-form single-PA placement and orientation \\
 & & \cmphl\cite{Ref_Bozkurt2026Trajectory} & \cmphl DL & \cmphl $2$--$10$ & \cmphl mobility, latency & \cmphl $1$\,s reconfiguration holds ${\sim}5$\,m/s, $0.5$\,s over $10$\,m/s; static loses 40\% at $5$\,m/s \\
\cmpgrp[12pt]{8}{Impairments} & & \cite{Ref_Li2026PatternMeasurement} & -- & -- & directivity & Prototype waveguide$+$PA; measured patterns of square/triangular PAs \\
\bottomrule
\end{tabular}
\\[2pt]
\parbox{\textwidth}{\raggedright\scriptsize
$^\ast$One physical waveguide but multiple modes/RF chains.\quad$^\dagger$Point-to-point, not multi-user.\\[1.5pt]
Acronyms: MAC, multiple-access channel; CNN, convolutional neural network; MEC, mobile-edge computing; CR, cognitive radio; QoS, quality of service; EE, energy efficiency; SE, spectral efficiency; KKT, Karush--Kuhn--Tucker; BS, base station; p2p, point-to-point.}
\end{table*}

\subsubsection{Single-Waveguide Families}
The single RF-fed waveguide is the field's most-studied and most-constrained setting, separating users only in time, frequency, power, or guided mode:
\begin{itemize}
\item \emph{Time-division multiple access (TDMA).} Far from a weak baseline, TDMA, if re-pinched per slot, is provably optimal in several settings: an alternating time-sharing scheme achieves the single-waveguide uplink sum capacity, making SIC unnecessary at the sum-rate vertex \cite{Ref_Chen2025SumCapacityMAC}; and, in wireless-powered edge computing, it matches NOMA under static activation but strictly dominates it once activation is fully dynamic \cite{Ref_Liu2025WPTMECTDMAvsNOMA}. It also outperforms NOMA at symmetric rate targets of $3$\,bit/s/Hz once four or more PAs are active \cite{Ref_Ren2025NOMAorOMA}. With the positions instead fixed per frame, it is max-min fair with a closed-form Lambert-$W$ allocation \cite{Ref_Tegos2025UplinkMinRate}. With a single PA, time division activates the antenna closest to each served user, whereas NOMA's fixed position penalizes the edge users farther from it \cite{Ref_Xie2025LowComplexityPlacement}. Its design is tractable at scale, a bipartite-graph formulation solved by a graph attention network holding the antenna positions fixed per frame---rather than re-pinching per slot---to curb reconfiguration overhead \cite{Ref_Xie2025BGATGNN}.
\item \emph{Frequency-division multiple access (FDMA).} Spatially separated PAs create a multi-tap channel; once the transmission bandwidth exceeds the resulting coherence bandwidth, the link turns frequency-selective and a single-carrier receiver suffers severe inter-symbol interference (ISI). FDMA assigns each user a disjoint sub-band; its multicarrier refinement, orthogonal FDMA (OFDMA), separates users by subcarrier and absorbs this selectivity via a cyclic prefix---the natural choice once transmission goes wideband \cite{Ref_Oikonomou2025OFDMA}.
\item \emph{Power-domain NOMA} is the dominant single-waveguide family, with placement engineering the channel-gain disparity that SIC needs. Its sub-problems are well developed: foundations and closed-form placement \cite{Ref_Ding2025PinchingPerspective,Ref_Ding2025AnalyticalPlacement}; discrete-grid antenna activation \cite{Ref_Wang2025AntennaActivationNOMA,Ref_Zhang2026EEDiscreteActivation}; power allocation \cite{Ref_Mohammadzadeh2025PowerAllocNOMA,Ref_Mohammadzadeh2025CNNPower,Ref_Vashakidze2025RadiationPower}; joint placement-and-power rate maximization \cite{Ref_Xie2025LowComplexityPlacement,Ref_Zhou2025SumRateNOMA,Ref_Xu2025QoSCRNOMA}; outage, ergodic-rate, and bit-error-rate (BER) analysis \cite{Ref_Cheng2025OMANOMAPerformance,Ref_Cheng2026SingleWGOMAvsNOMA,Ref_Yue2026NOMAPerformance,Ref_AlaaEldin2026BERNOMA}; uplink designs \cite{Ref_Zeng2025ULSumRateNOMA,Ref_Zeng2025ULEENOMA}; fairness, robustness, and security \cite{Ref_Ye2026FairnessNOMACSI,Ref_Chi2025NOMAEavesdropping}; and applications in edge computing, over-the-air computation, and semantics \cite{Ref_Ai2026NOMAMECDelay,Ref_Pakravan2026AirCompNOMA,Ref_Ahmed2026SemanticNOMA}. A partial variant tunes a continuous OMA$\leftrightarrow$NOMA overlap on the subcarriers, attaining the highest sum rate of the three across LoS and NLoS links, with the optimal pinching position migrating with the overlap \cite{Ref_Zhuo2026PNOMA}.
\item \emph{Mode-division multiplexing (MDM)} excites multiple orthogonal guided modes on one physical waveguide for independent streams---at the price of inter-mode power leakage through mismatched PAs (absorbed by zero-forcing) and one RF chain and feed per mode, so it relaxes the single-RF premise rather than living within it \cite{Ref_Xu2026MultiModeMDM}.
\item \emph{Segmented waveguide-enabled pinching-antenna systems (SWANs)} \cite{Ref_Ouyang2025SWAN} split the waveguide into per-feed segments, removing---via one active PA per segment---the uplink IAR that otherwise makes the shared-guide uplink intractable, and truncating the IWA a single long guide accumulates \cite{Ref_Gu2025SWANUplink}. Its segment-selection (SS) and segment-aggregation (SA) modes retain a single RF chain, whereas \emph{segment multiplexing} (SM) dedicates one RF chain per segment and thus reaches into the multi-waveguide regime (Fig.~\ref{fig:tax}), albeit as region-wise spatial reuse across co-linear segments rather than the co-located nulling of parallel waveguides; in performance, SM$>$SA$>$SS, with SA capturing most of the SM gain on a single RF chain.
\end{itemize}

\subsubsection{Multi-Waveguide Families}
The multi-waveguide regime spends one RF chain per guide to buy the spatial degrees of freedom a single guide cannot offer:
\begin{itemize}
\item \emph{SDMA/waveguide multiplexing (WM).} A baseband precoder across guides separates users in space---the multi-waveguide counterpart of multiple-input multiple-output (MIMO) precoding, jointly optimizing the precoding matrix and the activated pinching locations \cite{Ref_Bereyhi2025MIMO,Ref_MIMOPASS2025DLUL}; the waveguide-multiplexing/division/switching (WM/WD/WS) framework places SDMA alongside its lower-complexity cousins and gives the idealized, crosstalk-free rule that full precoding pays off most for co-located users \cite{Ref_Zhao2025TransmissionStructures}---a rule that residual impairments revise: in the benchmark, crosstalk removes most of the gain of full precoding over switching, and misalignment makes it pay off most for separated users (Section~\ref{ssec:e5}). Its uplink dual combines across guides rather than precoding into them: multiuser MISO with joint PA placement and user-power optimization \cite{Ref_Zhang2025ULMISO}.
\item \emph{Waveguide-division multiple access (WDMA) and environment-division multiple access (EDMA)} treat the waveguide itself \cite{Ref_Ding2025WDMA}---or the propagation environment via deliberate blockage \cite{Ref_Ding2025EDMA}---as the multiplexing resource; both are spatial-OMA-like and avoid joint baseband processing---WDMA across guides, EDMA across the segments of one guide---with an idealized downlink study comparing waveguide division directly against NOMA \cite{Ref_Zhang2026WDMAvsNOMA}. A power-domain overlay assigns users to guides and runs NOMA within each; the optimal per-guide SIC order is by interference-plus-noise-to-gain ratio and is power-allocation-invariant \cite{Ref_Wang2025MultiWGActNOMA}.
\item \emph{Multi-cell PASS coordination} manages inter-cell interference: coherent joint transmission converts it into useful signal \cite{Ref_Amhaz2025CoMP}, cell-free coordination counters the inter-cell interference and pilot contamination that cellular PASS aggravates \cite{Ref_Li2025CellFree}, a stochastic-geometry analysis shows pinching does not raise the network interference floor \cite{Ref_Sun2025MultiCellSG}, a Voronoi-topology analysis finds large-scale deployment evens the cell layout and lifts average rates \cite{Ref_Zhu2026Topology}, and ``whispering''---without data sharing or tight synchronization---lowers it by geometry alone \cite{Ref_Ding2026QuietWorld}.
\end{itemize}

\subsubsection{Cross-Regime and Emerging Families}
A few families resist the single-/multi-waveguide split, either interpolating between the two regimes or repurposing PASS-specific resources:
\begin{itemize}
\item \emph{Rate-splitting multiple access (RSMA)} is now an emerging family spanning single- and multi-waveguide and uplink and downlink settings \cite{Ref_RSMA2025MultiWG,Ref_RSMA2025Uplink}; by splitting messages into common and private parts it interpolates between SDMA and NOMA, generalizing power-domain access.
\item \emph{Index/spatial modulation} conveys bits through \emph{which} PA (or position) radiates; current schemes are point-to-point rather than multi-user, so they sit at the edge of the MA taxonomy \cite{Ref_Yang2025PAIM,Ref_PASM2026SpatialMux}.
\item \emph{Absent and nascent families.} Code-domain access---code-division multiple access (CDMA), sparse code multiple access (SCMA), and interleave-division multiple access (IDMA)---has, to our knowledge, no PASS instantiation to date, whereas grant-free/massive random access has only just been broached by a single SWAN-based contention protocol \cite{Ref_Shan2026SWANAccess}.
\end{itemize}

\begin{table}[!b]
\renewcommand{\arraystretch}{1.05}
\setlength{\tabcolsep}{5pt}
\DeclareRobustCommand{\cmark}{\scalebox{1.35}{\ding{51}}}%
\DeclareRobustCommand{\xmark}{\scalebox{0.85}{\ding{55}}}%
\DeclareRobustCommand{\pmark}{\tikz[baseline=-0.5ex]{\draw[line width=0.4pt,black!70] (0,0) circle (0.7ex);\fill[black!55] (0,0.7ex) arc (90:270:0.7ex) -- cycle;}}%
\caption{Existing PASS Surveys/Tutorials Versus This Article.\protect\\Marking: \cmark~yes; \pmark~partial; \xmark~no.}
\label{tab:surveys}
\centering
\footnotesize
\rowcolors{3}{acrzeb}{white}
\begin{tabular}{@{}p{2.6cm}p{2.85cm}cc@{}}
\toprule
\rowcolor{headblue}\textcolor{white}{\textbf{Survey}} & \textcolor{white}{\textbf{Primary focus}} & \textcolor{white}{\textbf{MA-}} & \textcolor{white}{\textbf{Impair.-}} \\
\rowcolor{headblue} & & \textcolor{white}{\textbf{centric}} & \textcolor{white}{\textbf{aware}} \\
\midrule
Yang \emph{et al.}\ \cite{Ref_Yang2025Principles} & General overview & \xmark & \xmark \\
Liu \emph{et al.}\ \cite{Ref_Liu2026Architecture} & Architecture/outlook & \xmark & \xmark \\
Liu \emph{et al.}\ \cite{Ref_Liu2025Tutorial} & Signal-model tutorial & \pmark & \pmark \\
Xu \emph{et al.}\ \cite{Ref_Xu2025GeneralizedPASS} & Media realizations & \pmark & \pmark \\
Zeng \emph{et al.}\ \cite{Ref_Zeng2025ResourceAlloc} & Resource allocation & \pmark & \xmark \\
Fang \emph{et al.}\ \cite{Ref_Fang2025MeetAI} & AI/ML for PASS & \xmark & \xmark \\
Liu \emph{et al.}\ \cite{Ref_Liu2026PASSSurvey} & Comprehensive survey & \xmark & \xmark \\
Illi and Qaraqe \cite{Ref_PASS2026ComprehensiveSurvey} & Comprehensive survey & \pmark & \xmark \\
Lu \emph{et al.}\ \cite{Ref_Lu2026OptimizationSurvey} & Optimization/AI designs & \pmark & \xmark \\
Zhao \emph{et al.}\ \cite{Ref_Zhao2026ReconfigAntennaSurvey} & Reconfigurable antennas & \xmark & \xmark \\
\midrule
\rowcolor{headblue!15}\textbf{This article} & \textbf{Multiple access} & \cmark & \cmark \\
\bottomrule
\end{tabular}
\end{table}

\subsection{The Idealization Gap, Scope, and Contributions}\label{ssec:gap}
Table~\ref{tab:cmp} summarizes the surveyed works and exposes the gap that motivates this article: of the 48 among them in the access families, 37 model none of the impairments within the scope of this article, and the remaining 11 model at most two, each in isolation---the prevailing model being \emph{idealized}: pure LoS propagation, perfect channel-state information (CSI) and SIC, negligible IWA, instantaneous PA repositioning, and omnidirectional radiation. Yet PASS is a \emph{mechanically reconfigured, electromagnetically coupled} system in which a single composite signal is shared across radiating points---a setting in which realistic impairments are not second-order corrections but first-order, method-specific effects. Mechanical reconfiguration---$20$--$30$\,ms per move, with update cadences of $0.5$--$1$\,s in practice \cite{Ref_Bozkurt2026Trajectory}---injects latency and timing jitter that strike time-division access directly; pinches on a shared guide couple electromagnetically \cite{Ref_Wang2025Modeling}, leaving the array gain non-monotone in their number and spacing \cite{Ref_Ouyang2025ArrayGain}; the guide attenuates at $0.08$--$1.3$\,dB/m, so IWA is negligible only over bounded regions \cite{Ref_Xu2025InWGAtten,Ref_Zhang2025DiPASS}; a single shared aperture radiates one frequency-dispersive, directionally selective pattern; and on the uplink a captured signal re-radiates through the other pinches. Electromagnetic modeling shows the ubiquitous omnidirectional assumption to be quantitatively wrong---omni-optimized placements lose about $7.5$\,dB \cite{Ref_Zubair2026Pattern}---a directional channel model has followed \cite{Ref_Zhang2025DiPASS}, and prototype-based pattern measurements confirm strongly geometry-dependent radiation \cite{Ref_Li2026PatternMeasurement}. Identifying \emph{which} impairment degrades \emph{which} access scheme, and by how much, is the gap this article fills.

Several valuable PASS overviews exist---on principles \cite{Ref_Yang2025Principles}, architecture and outlook \cite{Ref_Liu2026Architecture}, signal-model tutorials \cite{Ref_Liu2025Tutorial}, generalized media realizations \cite{Ref_Xu2025GeneralizedPASS}, resource allocation \cite{Ref_Zeng2025ResourceAlloc}, machine learning \cite{Ref_Fang2025MeetAI}, and two comprehensive 2026 surveys \cite{Ref_Liu2026PASSSurvey,Ref_PASS2026ComprehensiveSurvey}, alongside a survey of optimization, learning-based, and multifunctional designs \cite{Ref_Lu2026OptimizationSurvey} and a multi-technology reconfigurable-antenna survey that covers PASS \cite{Ref_Zhao2026ReconfigAntennaSurvey}---but none isolates multiple access as its organizing question, nor examines the access families from a realistic-impairment perspective (Table~\ref{tab:surveys}). Our scope is multiple access: beamforming, channel estimation, and hardware modeling enter only where they bear on the MA layer. Our contributions are:
\begin{itemize}
\item a \emph{comprehensive survey} of PASS multiple access---a unified taxonomy organized by the spatial-DoF spine, spanning the single-waveguide regime, the multi-waveguide regime, and the cross-regime and emerging families, together with the channel- and hardware-modeling studies that the impairment analysis later builds on (Section~\ref{sec:catalog});
\item an \emph{impairment-centric analysis} of PASS multiple access, the article's principal contribution: individual impairments have been analyzed before, but always in isolation \cite{Ref_Xu2025InWGAtten,Ref_Ouyang2025ArrayGain,Ref_AlaaEldin2026BERNOMA,Ref_Ye2026FairnessNOMACSI,Ref_Wang2025Modeling,Ref_Zubair2026Pattern,Ref_Zhang2025DiPASS,Ref_Bozkurt2026Trajectory}; here they are unified at the multiple-access layer, where each perturbs the effective channel, leaks interference across users' orthogonal resources, or---on the shared-guide uplink---couples the antennas structurally. An impairment-versus-method severity map charts the damage, and the first two are summarized by a channel perturbation $\Delta g$ and a leakage term $\varepsilon$ with per-domain models (Section~\ref{sec:impair});
\item a \emph{unified numerical benchmark} that compares time\nobreakdash-, frequency\nobreakdash-, power\nobreakdash-, and space-domain access under \emph{identical} channel and impairment models---to our knowledge the first common-footing, impairment-aware comparison in the field (Section~\ref{sec:bench}); and
\item a set of \emph{cross-cutting findings}: each access scheme is generally most vulnerable to the impairment that corrupts the very resource it uses to separate users, and rate splitting, which generalizes power-domain access, recovers most of the per-user fairness that full cancellation loses. They give no universal ranking of the schemes and point to the open problems (Section~\ref{sec:concl})---the untouched code-domain, the only-just-broached grant-free access, and the move from idealized analysis to hardware-validated study.
\end{itemize}

\begin{figure}[!b]
\centering
\begin{tikzpicture}[font=\sffamily\scriptsize,>=Stealth,
  sec/.style={rounded corners=2.5pt,draw=blue!55!black,line width=0.6pt,top color=blue!12,bottom color=blue!3,align=center,inner sep=3.5pt,text width=7.4cm},
  core/.style={rounded corners=2.5pt,draw=orange!80!black,line width=0.8pt,top color=orange!22,bottom color=orange!6,align=center,inner sep=3.5pt,text width=6.7cm},
  aux/.style={rounded corners=2.5pt,draw=black!45,line width=0.5pt,top color=black!8,bottom color=black!2,align=center,inner sep=3.5pt,text width=7.4cm},
  fl/.style={->,line width=0.9pt,draw=blue!50!black},
  hf/.style={->,line width=0.9pt,draw=orange!85!black}]
  \fill[rounded corners=4pt,orange!75!black,opacity=0.08] (-3.85,3.95) rectangle (3.85,7.35);
  \draw[rounded corners=4pt,draw=orange!70!black,line width=0.7pt,dash pattern=on 3pt off 2pt] (-3.85,3.95) rectangle (3.85,7.35);
  \node[anchor=north,font=\sffamily\scriptsize\bfseries,text=orange!58!black] at (0,7.27) {Impairment-aware core:\; \textsc{identify} $\to$ \textsc{analyze} $\to$ \textsc{evaluate}};
  \node[sec] (intro) at (0,9.4) {\textbf{Sec.~I: Introduction}\\ motivation \,\textbullet\, taxonomy of $56$ works (Fig.~\ref{fig:tax}) \,\textbullet\, idealization gap \,\textbullet\, scope \& contributions};
  \node[sec] (model) at (0,8.2) {\textbf{Sec.~II: System Model}\\ per-antenna power model \,\textbullet\, downlink / uplink / multi-waveguide signal models \,\textbullet\, structural penalties};
  \node[core] (identify) at (0,6.30) {\textbf{Sec.~III: \textsc{identify} $\to$ \textsc{analyze}}\\ catalog all impairments (Table~\ref{tab:impair}); the $\Delta g/\varepsilon$ framework picks the critical few; per-domain mechanisms and the impairment-versus-method map};
  \node[core] (evaluate) at (0,4.60) {\textbf{Sec.~IV: \textsc{evaluate}}\\ unified benchmark across the separation domains};
  \node[aux] (concl) at (0,2.75) {\textbf{Sec.~V: Conclusion}\\ findings \,\textbullet\, no universal ranking \,\textbullet\, open problems};
  \draw[fl] (intro) -- (model);
  \draw[fl] (model) -- (0,7.35);
  \draw[hf] (identify) -- (evaluate);
  \draw[fl] (0,3.95) -- (concl);
\end{tikzpicture}
\caption{Outline and logical flow of this article. }
\label{fig:outline}
\end{figure}
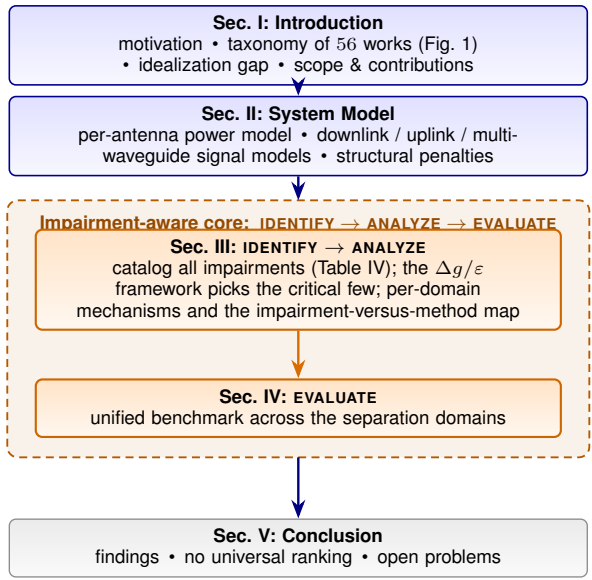

\subsection{Notation and Organization}
Boldface lowercase denotes vectors and boldface uppercase matrices; $\lVert\cdot\rVert$ is the Euclidean norm; $j=\sqrt{-1}$ is the imaginary unit; $\lambda$ is the free-space wavelength, $n_\eff$ the effective refractive index, $\lambda_g=\lambda/n_\eff$ the guided wavelength; $\sigma^2$ the noise power; $P$ the power budget; $N$ the number of active PAs on a guide; $K$ the number of users; $M$ the number of waveguides; and $\varepsilon$ denotes the orthogonality-leakage term, introduced in Section~\ref{sec:impair}. Table~\ref{tab:acronyms} lists the acronyms used throughout, and Fig.~\ref{fig:outline} maps the article's structure and logical flow. Section~\ref{sec:fund} develops the system model---the power model; the downlink, uplink, and multi-waveguide channels and architectures; and the structural penalties of a shared feed. The impairment-aware core then proceeds in two steps: Section~\ref{sec:impair} \emph{identifies} the multiple-access--critical impairments through the channel-perturbation/orthogonality-leakage framework and \emph{analyzes} them domain by domain; Section~\ref{sec:bench} then \emph{evaluates} them on a unified benchmark. Section~\ref{sec:concl} concludes with the findings, what they imply, and the open problems.

\section{System Model}\label{sec:fund}\label{sec:model}
This section builds the signal model used throughout the article. It covers how the transmit power is divided among the PAs; the downlink, uplink, and multi-waveguide channels; and the three penalties a shared feed imposes---power dilution and the phase-alignment conflict on the downlink, and IAR on the uplink.

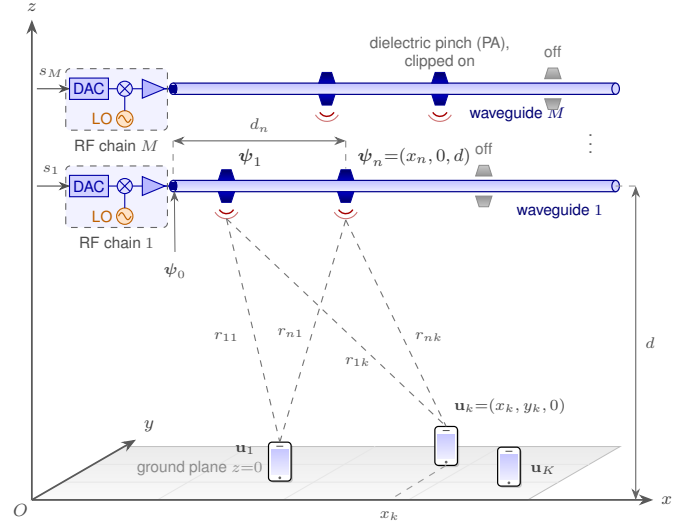
\begin{figure}[!t]
\centering
\resizebox{\columnwidth}{!}{%
\begin{tikzpicture}[>=Stealth,font=\sffamily\footnotesize,line width=0.5pt]
  \shade[top color=black!8, bottom color=black!2] (0,0) -- (7.9,0) -- (9.37,0.85) -- (1.47,0.85) -- cycle;
  \draw[black!25, line width=0.3pt] (0,0) -- (7.9,0) -- (9.37,0.85) -- (1.47,0.85) -- cycle;
  \draw[black!10, line width=0.3pt] (2,0) -- (3.47,0.85) (4,0) -- (5.47,0.85) (6,0) -- (7.47,0.85);
  \draw[black!10, line width=0.3pt] (0.49,0.283) -- (8.39,0.283) (0.98,0.567) -- (8.88,0.567);
  \draw[->,black!65,thick] (0,0) -- (9.9,0) node[right,text=black!75] {$x$};
  \draw[->,black!65,thick] (0,0) -- (1.66,0.96) node[above right,text=black!75] {$y$};
  \draw[->,black!65,thick] (0,0) -- (0,7.65) node[above,text=black!75] {$z$};
  \node[anchor=north east,inner sep=1.5pt,text=black!70] at (0,0) {$O$};
  \node[black!45,anchor=west,font=\sffamily\scriptsize] at (1.55,0.50) {ground plane $z{=}0$};
  \shade[top color=blue!24, bottom color=blue!6] (2.25,6.46) rectangle (9.30,6.64);
  \draw[blue!55!black,line width=0.55pt] (2.25,6.46) -- (9.30,6.46) (2.25,6.64) -- (9.30,6.64);
  \draw[blue!55!black,fill=blue!16,line width=0.55pt] (9.30,6.55) ellipse (0.06 and 0.09);
  \draw[blue!55!black,fill=blue!16,line width=0.55pt] (2.25,6.55) ellipse (0.06 and 0.09);
  \node[blue!45!black,anchor=north west,font=\sffamily\scriptsize] at (6.85,6.38) {waveguide $M$};
  \foreach \px in {4.7,6.5} {
    \shade[top color=blue!40!black, bottom color=blue!75!black] (\px-0.13,6.64) -- (\px+0.13,6.64) -- (\px+0.075,6.81) -- (\px-0.075,6.81) -- cycle;
    \shade[top color=blue!75!black, bottom color=blue!40!black] (\px-0.13,6.46) -- (\px+0.13,6.46) -- (\px+0.075,6.29) -- (\px-0.075,6.29) -- cycle;
    \draw[blue!60!black,line width=0.4pt] (\px-0.13,6.64) -- (\px+0.13,6.64) -- (\px+0.075,6.81) -- (\px-0.075,6.81) -- cycle;
    \draw[blue!60!black,line width=0.4pt] (\px-0.13,6.46) -- (\px+0.13,6.46) -- (\px+0.075,6.29) -- (\px-0.075,6.29) -- cycle;
    \draw[red!65!black,line width=0.65pt] ($(\px,6.24)+(210:0.13)$) arc (210:330:0.13);
    \draw[red!65!black!55,line width=0.5pt] ($(\px,6.24)+(210:0.21)$) arc (210:330:0.21);
  }
  \shade[top color=black!22, bottom color=black!42] (8.3-0.13,6.71) -- (8.3+0.13,6.71) -- (8.3+0.075,6.88) -- (8.3-0.075,6.88) -- cycle;
  \shade[top color=black!42, bottom color=black!22] (8.3-0.13,6.39) -- (8.3+0.13,6.39) -- (8.3+0.075,6.22) -- (8.3-0.075,6.22) -- cycle;
  \draw[black!45,line width=0.4pt] (8.3-0.13,6.71) -- (8.3+0.13,6.71) -- (8.3+0.075,6.88) -- (8.3-0.075,6.88) -- cycle;
  \draw[black!45,line width=0.4pt] (8.3-0.13,6.39) -- (8.3+0.13,6.39) -- (8.3+0.075,6.22) -- (8.3-0.075,6.22) -- cycle;
  \node[anchor=south,font=\sffamily\scriptsize,text=black!55] at (8.3,6.93) {off};
  \node[anchor=north,align=center,font=\sffamily\scriptsize,text=black!60] at (6.5,7.50) {dielectric pinch (PA),\\clipped on};
  \fill[rounded corners=2pt,blue!5] (0.54,5.89) rectangle (2.16,6.87);
  \draw[->,black!70] (0.08,6.55) -- (0.61,6.55);
  \node[anchor=south,font=\sffamily\scriptsize,text=black!70] at (0.31,6.61) {$s_M$};
  \filldraw[fill=blue!18,draw=blue!65!black] (0.60,6.38) rectangle (1.22,6.72); \node[font=\sffamily\scriptsize,text=blue!65!black] at (0.91,6.55) {DAC};
  \draw[blue!65!black] (1.22,6.55) -- (1.36,6.55);
  \filldraw[fill=blue!12,draw=blue!65!black] (1.48,6.55) circle (0.12);
  \draw[blue!65!black] (1.395,6.465) -- (1.565,6.635); \draw[blue!65!black] (1.395,6.635) -- (1.565,6.465);
  \filldraw[fill=orange!22,draw=orange!75!black] (1.48,6.07) circle (0.13);
  \draw[orange!75!black,line width=0.55pt] (1.38,6.07) sin (1.43,6.125) cos (1.48,6.07) sin (1.53,6.015) cos (1.58,6.07);
  \node[font=\sffamily\scriptsize,text=orange!75!black,anchor=east,inner sep=1pt] at (1.33,6.07) {LO};
  \draw[blue!65!black] (1.48,6.20) -- (1.48,6.43);
  \draw[blue!65!black] (1.60,6.55) -- (1.76,6.55);
  \filldraw[fill=blue!28,draw=blue!65!black] (1.76,6.40) -- (1.76,6.70) -- (2.12,6.55) -- cycle;
  \draw[blue!65!black,line width=0.7pt] (2.12,6.55) -- (2.21,6.55);
  \fill[blue!55!black] (2.21,6.49) rectangle (2.31,6.61);
  \draw[dashed,rounded corners=2pt,black!60] (0.54,5.89) rectangle (2.16,6.87);
  \node[anchor=north,font=\sffamily\scriptsize,text=black!70] at (1.35,5.86) {RF chain $M$};
  \node[text=black!60] at (8.9,5.78) {$\vdots$};
  \shade[top color=blue!24, bottom color=blue!6] (2.25,4.91) rectangle (9.30,5.09);
  \draw[blue!55!black,line width=0.55pt] (2.25,4.91) -- (9.30,4.91) (2.25,5.09) -- (9.30,5.09);
  \draw[blue!55!black,fill=blue!16,line width=0.55pt] (9.30,5.0) ellipse (0.06 and 0.09);
  \draw[blue!55!black,fill=blue!16,line width=0.55pt] (2.25,5.0) ellipse (0.06 and 0.09);
  \node[blue!45!black,anchor=north west,font=\sffamily\scriptsize] at (7.6,4.84) {waveguide $1$};
  \fill[rounded corners=2pt,blue!5] (0.54,4.34) rectangle (2.16,5.32);
  \draw[->,black!70] (0.08,5.0) -- (0.61,5.0);
  \node[anchor=south,font=\sffamily\scriptsize,text=black!70] at (0.31,5.06) {$s_1$};
  \filldraw[fill=blue!18,draw=blue!65!black] (0.60,4.83) rectangle (1.22,5.17); \node[font=\sffamily\scriptsize,text=blue!65!black] at (0.91,5.0) {DAC};
  \draw[blue!65!black] (1.22,5.0) -- (1.36,5.0);
  \filldraw[fill=blue!12,draw=blue!65!black] (1.48,5.0) circle (0.12);
  \draw[blue!65!black] (1.395,4.915) -- (1.565,5.085); \draw[blue!65!black] (1.395,5.085) -- (1.565,4.915);
  \filldraw[fill=orange!22,draw=orange!75!black] (1.48,4.52) circle (0.13);
  \draw[orange!75!black,line width=0.55pt] (1.38,4.52) sin (1.43,4.575) cos (1.48,4.52) sin (1.53,4.465) cos (1.58,4.52);
  \node[font=\sffamily\scriptsize,text=orange!75!black,anchor=east,inner sep=1pt] at (1.33,4.52) {LO};
  \draw[blue!65!black] (1.48,4.65) -- (1.48,4.88);
  \draw[blue!65!black] (1.60,5.0) -- (1.76,5.0);
  \filldraw[fill=blue!28,draw=blue!65!black] (1.76,4.85) -- (1.76,5.15) -- (2.12,5.0) -- cycle;
  \draw[blue!65!black,line width=0.7pt] (2.12,5.0) -- (2.21,5.0);
  \fill[blue!55!black] (2.21,4.94) rectangle (2.31,5.06);
  \node[anchor=north,font=\sffamily\scriptsize,text=black!70] at (2.28,3.90) {$\boldsymbol{\psi}_0$};
  \draw[->,black!60,line width=0.4pt] (2.28,3.95) -- (2.27,4.92);
  \draw[dashed,rounded corners=2pt,black!60] (0.54,4.34) rectangle (2.16,5.32);
  \node[anchor=north,font=\sffamily\scriptsize,text=black!70] at (1.35,4.31) {RF chain $1$};
  \foreach \px in {3.1,5.0} {
    \shade[top color=blue!40!black, bottom color=blue!75!black] (\px-0.13,5.09) -- (\px+0.13,5.09) -- (\px+0.075,5.26) -- (\px-0.075,5.26) -- cycle;
    \shade[top color=blue!75!black, bottom color=blue!40!black] (\px-0.13,4.91) -- (\px+0.13,4.91) -- (\px+0.075,4.74) -- (\px-0.075,4.74) -- cycle;
    \draw[blue!60!black,line width=0.4pt] (\px-0.13,5.09) -- (\px+0.13,5.09) -- (\px+0.075,5.26) -- (\px-0.075,5.26) -- cycle;
    \draw[blue!60!black,line width=0.4pt] (\px-0.13,4.91) -- (\px+0.13,4.91) -- (\px+0.075,4.74) -- (\px-0.075,4.74) -- cycle;
    \draw[red!65!black,line width=0.65pt] ($(\px,4.69)+(210:0.13)$) arc (210:330:0.13);
    \draw[red!65!black!55,line width=0.5pt] ($(\px,4.69)+(210:0.21)$) arc (210:330:0.21);
  }
  \shade[top color=black!22, bottom color=black!42] (7.2-0.13,5.16) -- (7.2+0.13,5.16) -- (7.2+0.075,5.33) -- (7.2-0.075,5.33) -- cycle;
  \shade[top color=black!42, bottom color=black!22] (7.2-0.13,4.84) -- (7.2+0.13,4.84) -- (7.2+0.075,4.67) -- (7.2-0.075,4.67) -- cycle;
  \draw[black!45,line width=0.4pt] (7.2-0.13,5.16) -- (7.2+0.13,5.16) -- (7.2+0.075,5.33) -- (7.2-0.075,5.33) -- cycle;
  \draw[black!45,line width=0.4pt] (7.2-0.13,4.84) -- (7.2+0.13,4.84) -- (7.2+0.075,4.67) -- (7.2-0.075,4.67) -- cycle;
  \node[anchor=south,font=\sffamily\scriptsize,text=black!55] at (7.2,5.38) {off};
  \node[anchor=south west,text=black!80,inner sep=1pt] at (3.25,5.28) {$\boldsymbol{\psi}_1$};
  \node[anchor=south west,text=black!80,inner sep=1pt] at (5.15,5.28) {$\boldsymbol{\psi}_n{=}(x_n,0,d)$};
  \draw[<->,black!60] (2.25,5.72) -- (5.0,5.72);
  \draw[dashed,black!45] (2.25,5.72) -- (2.25,5.11);
  \draw[dashed,black!45] (5.0,5.72) -- (5.0,5.28);
  \node[anchor=south,font=\sffamily\scriptsize,text=black!70] at (3.63,5.74) {$d_n$};
  \foreach \px/\py/\nm in {3.95/0.30/u1, 6.6/0.55/u2, 7.6/0.22/u3} {
    \draw[fill=white,rounded corners=1.6pt,line width=0.55pt,
          drop shadow={shadow xshift=0.8pt, shadow yshift=-0.8pt, fill=black, opacity=0.18}]
          (\px-0.18,\py) rectangle (\px+0.18,\py+0.62);
    \shade[top color=blue!22, bottom color=blue!6] (\px-0.135,\py+0.105) rectangle (\px+0.135,\py+0.50);
    \draw[line width=0.45pt,black!70] (\px-0.055,\py+0.555) -- (\px+0.055,\py+0.555);
    \fill[black!60] (\px,\py+0.052) circle (0.022);
    \coordinate (\nm) at (\px,\py+0.62);
  }
  \node[anchor=east,font=\sffamily\scriptsize,text=black!75] at (3.68,0.80) {$\mathbf{u}_1$};
  \node[anchor=south west,font=\sffamily\scriptsize,text=black!75] at (6.60,1.25) {$\mathbf{u}_k{=}(x_k,y_k,0)$};
  \draw[dashed,black!40] (6.6,0.55) -- (5.7,0);
  \node[anchor=north,font=\sffamily\scriptsize,text=black!70] at (5.7,-0.05) {$x_k$};
  \node[anchor=west,font=\sffamily\scriptsize,text=black!75] at (7.82,0.48) {$\mathbf{u}_K$};
  \draw[dashed,black!55] (3.1,4.48) -- (u1); \node[anchor=east,font=\sffamily\scriptsize,text=black!70] at (3.42,2.60) {$r_{11}$};
  \draw[dashed,black!55] (3.1,4.48) -- (u2); \node[anchor=east,font=\sffamily\scriptsize,text=black!70] at (5.52,2.20) {$r_{1k}$};
  \draw[dashed,black!55] (5.0,4.46) -- (u1); \node[anchor=east,font=\sffamily\scriptsize,text=black!70] at (4.46,2.66) {$r_{n1}$};
  \draw[dashed,black!55] (5.0,4.46) -- (u2); \node[anchor=west,font=\sffamily\scriptsize,text=black!70] at (5.98,2.60) {$r_{nk}$};
  \draw[dashed,black!40] (9.30,5.0) -- (9.62,5.0);
  \draw[<->,black!60] (9.62,0) -- (9.62,5.0);
  \node[anchor=west,font=\sffamily\scriptsize,text=black!70] at (9.65,2.50) {$d$};
\end{tikzpicture}}
\caption{System model of a multi-waveguide pinching-antenna system. Each waveguide $m$ is fed by its own RF chain---digital-to-analog conversion, up-conversion by a local oscillator (LO), and amplification---injecting the composite signal $s_m$. The figure depicts the downlink; on the uplink the chain operates in receive and the same feed observes the back-propagated signal.}
\label{fig:sysmodel}
\end{figure}

A PASS feeds a low-loss dielectric waveguide from one end; small dielectric \emph{pinches} placed on the guide couple part of the guided power to free space, each acting as an antenna whose position can be added, removed, or repositioned on demand---the \emph{programmable} radiating geometry that distinguishes PASS and underlies pinching beamforming (Fig.~\ref{fig:sysmodel}). Coupled-mode theory gives the amplitude leaked by the $n$-th PA as a cascade along the guide,
\begin{equation}
s_{\mathrm{rad},n}=\delta_n\!\prod_{i=1}^{n-1}\!\sqrt{1-\delta_i^2}\;e^{-j\frac{2\pi}{\lambda_g}d_n},\qquad \delta_n=\sin(\kappa L_n),
\label{eq:cmt}
\end{equation}
where $\kappa$ is the coupling coefficient, $L_n$ the coupling length of PA $n$, and $d_n$ its distance from the feed along the guide. Two power conventions follow:
\begin{enumerate}
\item \emph{Equal power} ($P_n=P/N$): every PA radiates the same power. Since each pinch depletes the guide, this needs a different coupling length $L_n$ per PA, growing along the guide---so the PAs are not interchangeable. This is the analytically convenient default of most MA studies.
\item \emph{Proportional power}: all PAs are \emph{identical} ($L_n=L$), so each radiates the same \emph{fraction} of the power still travelling in the guide. The radiated amplitudes then decay geometrically, $s_{\mathrm{rad},n}\propto\delta(\sqrt{1-\delta^2})^{\,n-1}$, but the PAs are interchangeable---a single pinch design serves every position.
\end{enumerate}
The two nearly coincide for short, low-loss guides with few PAs, so most small-$N$ MA results are insensitive to the choice \cite{Ref_Wang2025Modeling}; the gap grows with $N$ and IWA, so the distinction---and the IWA neglected in \eqref{eq:cmt}---matters for large active sets, long guides \cite{Ref_Xu2025InWGAtten}, and terahertz operation \cite{Ref_Jiang2026THzPASS}. The PA positions $\{\boldsymbol{\psi}_n\}$ are reconfigured by one of three mechanisms \cite{Ref_Liu2025Tutorial}:
\begin{enumerate}
\item \emph{Continuous} activation slides a PA along a track via a servo or stepper (milliseconds to seconds, fine).
\item \emph{Discrete} activation switches PAs pre-installed at fixed candidate slots on and off (microseconds to milliseconds, coarse).
\item \emph{Semi-continuous} activation refines a discrete grid with a local micro-actuator.
\end{enumerate}
Discrete activation is attractive for cost and speed but needs a dense grid---hundreds of candidate positions per meter at millimeter-wave frequencies---to approach continuous performance, because the in-waveguide phase $2\pi d_n/\lambda_g$ must be sampled finely. The reconfiguration timescale and grid resolution are precisely the hardware parameters through which the latency and the position error of Section~\ref{sec:impair} act.

\subsection{Downlink Signal Model}\label{ssec:singlewg}
Consider first a \emph{single} waveguide along the $x$-axis at height $d$ (Fig.~\ref{fig:sysmodel}, waveguide 1), fed by a single RF chain at $\boldsymbol{\psi}_0$, with $N$ active PAs at $\{\boldsymbol{\psi}_n\}_{n=1}^N$ serving single-antenna users $\{\mathbf{u}_k\}_{k=1}^{K}$. The guided wave accumulates the in-waveguide phase $2\pi d_n/\lambda_g$ over the feed-to-PA distance $d_n=\lVert\boldsymbol{\psi}_0-\boldsymbol{\psi}_n\rVert$, with $\lambda_g=\lambda/n_\eff$; the free-space distance from PA $n$ to user $k$ is $r_{nk}=\lVert\mathbf{u}_k-\boldsymbol{\psi}_n\rVert$. The signal received by user $k$ is given by
\begin{equation}
y_k=\sum_{n=1}^{N}\frac{c\sqrt{P_n}}{4\pi f_c\,r_{nk}}\,e^{-j\frac{2\pi}{\lambda}r_{nk}}\,e^{-j\frac{2\pi}{\lambda_g}d_n}\, s+w_k,
\end{equation}
where $c$ is the speed of light, $f_c$ the carrier frequency, $P_n$ the power radiated by PA $n$ (set passively by its coupling length [cf.~\eqref{eq:cmt}]), $s$ a unit-power composite symbol, and $w_k\sim\mathcal{CN}(0,\sigma^2)$. Under the equal-power convention $P_n=P/N$, user $k$'s \emph{effective channel} is
\begin{equation}
g_k=\sum_{n=1}^{N}\frac{c}{4\pi f_c\,r_{nk}}\exp\!\Big[-j2\pi\Big(\tfrac{r_{nk}}{\lambda}+\tfrac{d_n}{\lambda_g}\Big)\Big],
\label{eq:effchan}
\end{equation}
so that $y_k=\sqrt{P/N}\,g_k\,s+w_k$. Equation~\eqref{eq:effchan} is central to PASS: displacing one PA changes its amplitude $1/r_{nk}$, its free-space phase $2\pi r_{nk}/\lambda$, and its in-waveguide phase $2\pi d_n/\lambda_g$ at once \cite{Ref_Ding2025PinchingPerspective}. The objective over positions therefore oscillates at the guided-wavelength scale ($\lambda_g\approx7.65$\,mm at $28$\,GHz with $n_\eff\approx1.4$), producing a dense set of local optima; pinching beamforming is far more capable than RIS phase shifting---it reshapes large-scale path loss, not just phase---but markedly harder to optimize and acutely sensitive to position error \cite{Ref_AlaaEldin2026BERNOMA} (the hardware root of the phase-alignment conflict; see Remark~\ref{rem:phase}). User $k$'s signal-to-noise ratio (SNR) is $\gamma_k=(P/\sigma^2)\,a(N)$ with the per-antenna-normalized array gain
\begin{equation}
a(N)=\frac1N\left|\sum_{n=1}^N \tfrac{c}{4\pi f_c r_{nk}}\,e^{-j2\pi(\frac{r_{nk}}{\lambda}+\frac{d_n}{\lambda_g})}\right|^2.
\label{eq:arraygain}
\end{equation}
\begin{remark}[Power dilution]\label{rem:dilution}
The budget splits across the $N$ PAs, so \eqref{eq:arraygain} carries the prefactor $1/N$. Adding PAs both inserts terms into the sum and shrinks the prefactor; since distant PAs contribute little magnitude, $a(N)$ is generally \emph{non-monotone}, with a finite optimum $N^\star$ \cite{Ref_Ouyang2025ArrayGain}---unlike a conventional array, where each element has its own power and ``more is better.'' Charging per-PA activation power sharpens this into a hard optimum. The same conservation argument binds across \emph{users}: a passive pinch radiates the whole composite signal, on all subcarriers alike, so the pinches nearest to the $K$ users share the power, and the share of the worst-served user cannot exceed $1/K$ (about $M/K$ with $M$ guides).
\end{remark}
\begin{remark}[Phase-alignment conflict]\label{rem:phase}
Constructive combining at a target user $\mathbf{u}_k$ requires
\begin{equation}
\phi_{nk}\triangleq\tfrac{2\pi}{\lambda}r_{nk}+\tfrac{2\pi}{\lambda_g}d_n\equiv \bar{\phi}_k\ (\mathrm{mod}\ 2\pi),\quad\forall n,
\label{eq:phase}
\end{equation}
for a common reference phase $\bar{\phi}_k$; the same condition for a second user $\mathbf{u}_{k'}$ involves $r_{nk'}\neq r_{nk}$ and is generally \emph{incompatible} with \eqref{eq:phase} on the same positions ($d_n$) once $N\ge2$ and $K\ge2$: a single waveguide radiates one phase profile and cannot be co-aligned at many users at once, and by reciprocity the same conflict governs uplink combining at the feed.
\end{remark}

\subsection{Uplink Signal Model}
By reciprocity, an active PA captures the user's signal and the guided wave travels back to the feed through the same coupled-mode cascade as in \eqref{eq:cmt}. With several PAs on one guide, however, part of that returning wave is re-radiated by every pinch it passes---\emph{inter-antenna radiation}. No tractable end-to-end multi-PA uplink model exists yet; the multiport-network PASS study leaves it explicitly open \cite{Ref_Wang2025Multiport}.
\begin{remark}[Inter-antenna radiation]\label{rem:iar}
Inter-antenna radiation is the uplink's structural penalty, the receive-side counterpart of the downlink's power dilution (Remark~\ref{rem:dilution}): unique to the shared guide, it electromagnetically couples every active PA to the others and is the obstacle that has motivated single-active-PA operation or segmentation.
\end{remark}
Existing uplink models fall into two groups. Some keep several PAs active but describe the pinch as a lossless phase shifter rather than a coupler, neglecting re-radiation altogether \cite{Ref_Chen2025SumCapacityMAC,Ref_Tegos2025UplinkMinRate}; others activate a single PA per slot or per segment, so no pinch is left to re-radiate---the standard way around the problem \cite{Ref_Ouyang2025SWAN,Ref_Gu2025SWANUplink}. We call the second group \emph{IAR-free} hereinafter; the first trades physical consistency for tractability. This paper is instead \emph{IAR-aware}: carrying the coupled-mode description of \eqref{eq:cmt} over to the uplink, the returning wave is depleted, to first order, at every pinch it passes. Writing $n'$ for the PA serving user $k$, the feed observes
\begin{equation}
\tilde{y}=\sum_{k=1}^{K}\sqrt{P_k}\,g_k^{\mathrm{ul}}\,x_k+w,
\end{equation}
where $P_k$ is user $k$'s transmit power, $x_k$ its unit-power symbol, $w\sim\mathcal{CN}(0,\sigma^2)$ the noise at the feed, and
\begin{equation}
g_k^{\mathrm{ul}}=\frac{c\,\delta_{n'}\prod_{i<n'}\sqrt{1-\delta_i^{2}}}{4\pi f_c\,r_{n'k}}\,e^{-j\big(\frac{2\pi}{\lambda}r_{n'k}+\frac{2\pi}{\lambda_g}d_{n'}\big)},
\label{eq:ul}
\end{equation}
with $\delta_{n'}$ of \eqref{eq:cmt} and the two distances $d_{n'}$ and $r_{n'k}$ of Section~\ref{ssec:singlewg} taken at the serving PA $n'$, which varies with $k$. Two limits place \eqref{eq:ul} against the literature: setting $\delta_{n'}\prod_{i<n'}\sqrt{1-\delta_i^{2}}\!=\!1$ discards the cascade and returns the IAR-neglecting transfer of \cite{Ref_Chen2025SumCapacityMAC,Ref_Tegos2025UplinkMinRate}, while a single active PA per slot or per segment empties the product and makes \eqref{eq:ul} exact---the IAR-free case of \cite{Ref_Ouyang2025SWAN,Ref_Gu2025SWANUplink}.

Downlink and uplink alike, the single RF chain is the binding constraint:
\begin{quote}
\emph{one waveguide $=$ one feed $=$ one RF chain $=$ one composite signal $s$.}
\end{quote}
On the downlink every PA radiates the same $s$, so there is no per-user precoder and hence \emph{no spatial-domain multiple access}; on the uplink the guide exposes one shared feed, i.e., a multiple-access channel subject to IAR (Remark~\ref{rem:iar}). Users sharing a guide must therefore be separated in time, frequency, or power, or in guided mode, which needs one RF chain per mode.

\subsection{Multi-Waveguide Signal Model}
These relations are the $M{=}1$ instance of a general system with $M$ waveguides, which changes the picture qualitatively. Let guide $m$ carry composite symbol $s_m$ and expose effective channel $g_{k,m}$ to user $k$ (evaluated as in \eqref{eq:effchan} for that guide's PAs and scaled by $1/\sqrt{N}$ for their equal power split). Stacking $\mathbf{g}_k=[g_{k,1},\dots,g_{k,M}]^{\!\top}$ and $\mathbf{s}=[s_1,\dots,s_M]^{\!\top}$,
\begin{equation}
y_k=\mathbf{g}_k^{\!\top}\mathbf{s}+w_k,\qquad k=1,\dots,K.
\label{eq:multiwg}
\end{equation}
Because the $M$ feeds can carry \emph{independent} streams, a baseband precoder $\mathbf{W}\in\mathbb{C}^{M\times K}$ can map the users' unit-power data symbols $\mathbf{x}=[x_1,\dots,x_K]^{\!\top}$ onto the feeds as $\mathbf{s}=\mathbf{W}\mathbf{x}$, so spatial multiplexing is recovered---impossible at $M{=}1$, where $\mathbf{s}$ collapses to the scalar $s$. Three transmission architectures use the $M$ guides at different cost \cite{Ref_Liu2025Tutorial,Ref_Zhao2025TransmissionStructures}:
\begin{itemize}
\item \emph{Waveguide switching}: one guide active at a time---the spatial analogue of TDMA.
\item \emph{Waveguide division}: one dedicated stream per guide with no cross-guide baseband processing; users (or clusters) are separated by which guide serves them---the basis of WDMA \cite{Ref_Ding2025WDMA}.
\item \emph{Waveguide multiplexing}: a full baseband precoder $\mathbf{W}$ across all guides, i.e., genuine SDMA; with the analog/phase-shifter and pinching layers it yields \emph{tri-hybrid} beamforming (digital, analog, and pinching).
\end{itemize}
The waveguide-to-user ratio thus sets the available spatial degrees of freedom (Fig.~\ref{fig:tax}): one RF-fed guide forbids SDMA, any $M\ge2$ guides admit spatial separation (with one stream per user once $M\ge K$; the intermediate $1<M<K$ regime hosts clustered hybrids that run NOMA within each guide \cite{Ref_Wang2025MultiWGActNOMA})---at the cost of the inter-waveguide coupling and multi-chain synchronization analyzed in Section~\ref{sec:impair}.

\section{Analysis of the Multiple-Access--Critical Impairments}\label{sec:impair}
Table~\ref{tab:impair} is deliberately exhaustive: it catalogs, in five physical classes, \emph{every} impairment a realistic PASS link can suffer---most of which are not specific to multiple access. The task of this section is to \emph{identify}, from this long list, the few impairments that can actually reshape the multiple-access behavior. Section~\ref{ssec:lens} provides the framework: it writes the data rate of every access family in a unified form, and an impairment is multiple-access critical only if it impacts that rate---by perturbing the effective channel ($\Delta g$), by leaking interference across users' orthogonal resources ($\varepsilon$), or, on the shared-guide uplink, by coupling the antennas structurally. These three categories also set the order of Section~\ref{ssec:perdomain}: $\Delta g$ first, since it perturbs the channel for every scheme; then the $\varepsilon$ leakages, each subsection with an access family that uses that domain; and last IAR. Each subsection gives the physical mechanism and its qualitative severity, and Section~\ref{ssec:meta} summarizes them in the impairment-versus-method map.

\begin{table}[!t]
\caption{Catalog of PASS Impairments. Their multiple-access role and the $\Delta g/\varepsilon$ selection are developed in this section.}
\label{tab:impair}
\centering
\footnotesize
\setlength{\tabcolsep}{4pt}
\renewcommand{\arraystretch}{1.1}
\rowcolors{2}{acrzeb}{white}
\begin{tabular}{@{}>{\raggedright\arraybackslash\bfseries}p{2.45cm}>{\raggedright\arraybackslash}p{4.35cm}>{\raggedright\arraybackslash}p{1.15cm}@{}}
\toprule
\rowcolor{headblue}\textcolor{white}{\textbf{Class}} & \textcolor{white}{\textbf{Impairment (mechanism)}} & \textcolor{white}{\textbf{MA role}} \\
\midrule
Actuation / reconfig. &
\textbullet~latency\newline
\textbullet~timing jitter\newline
\textbullet~grid quantization\newline
\textbullet~activation energy & time / $\Delta g$ \\
Placement / geometry &
\textbullet~PA position error\newline
\textbullet~mutual coupling & $\Delta g$ \\
Waveguide / EM &
\textbullet~in-waveguide attenuation\newline
\textbullet~dispersion \& beam squint\newline
\textbullet~inter-antenna radiation\newline
\textbullet~directional radiation pattern\newline
\textbullet~coupling-length / fabrication error\newline
\textbullet~impedance mismatch\newline
\textbullet~inter-waveguide crosstalk\newline
\textbullet~mode / polarization crosstalk & freq. / space / mode / UL \\
RF chain / Tx &
\textbullet~power amplifier nonlinearity \newline
\textbullet~peak-to-average power ratio\newline
\textbullet~phase noise\newline
\textbullet~ADC/DAC quantization\newline
\textbullet~multi-chain synchronization & mixed \\
Signal proc. / CSI &
\textbullet~imperfect SIC\newline
\textbullet~CSI error / feedback latency & power / $\Delta g$ \\
\bottomrule
\end{tabular}
\end{table}

\subsection{Unified Analytical Framework}\label{ssec:lens}
Every access family of the taxonomy (Section~\ref{sec:catalog}) writes user~$k$'s rate in the single form
\begin{equation}
R_k=\log_2(1+\gamma_k),\qquad \SINR_k=\frac{|g_k|^2 P_k}{\sigma^2+I_k},
\label{eq:sinr}
\end{equation}
where $g_k$ is user $k$'s effective channel---the single-guide, equal-power instance of \eqref{eq:effchan} evaluated at $\mathbf{u}_k$, the $1/\sqrt{N}$ power-dilution amplitude (Remark~\ref{rem:dilution}) absorbed---$P_k$ the power assigned to user $k$, and $I_k$ the residual inter-user interference that the scheme's separation mechanism is meant to drive toward zero. The four families differ only in their choices of the signal power $P_k$ and the residual $I_k$, as follows.
\begin{enumerate}
\item[(i)] \emph{OMA} allots orthogonal fractions $\tau_k$ ($\sum_k\tau_k\le1$) and keeps $I_k=0$ on each dedicated resource, which gives
\begin{equation}
R_k=\tau_k\log_2\!\left(1+\frac{|g_k|^2P}{\sigma^2}\right).
\end{equation}
Under time division, it re-pinches $\{\boldsymbol{\psi}_n\}$ per slot for the active user.
\item[(ii)] \emph{Power-domain NOMA} superposes users and drives $I_k$ down by SIC; for a strong/weak pair with effective channels $g_s,g_w$ ($|g_s|^2\ge|g_w|^2$) and split $\alpha_w+\alpha_s=1$, the data rates of the weak and the strong user are given by
\begin{align}
R_w&=\log_2\!\left(1+\frac{\alpha_w|g_w|^2P}{\alpha_s|g_w|^2P+\sigma^2}\right),\label{eq:nomaw}\\
R_s&=\log_2\!\left(1+\frac{\alpha_s|g_s|^2P}{\sigma^2}\right),\label{eq:nomas}
\end{align}
with PASS \emph{engineering} the gain ordering through placement.
\item[(iii)] \emph{SDMA} (multi-waveguide) precodes $\mathbf{s}=\sum_k\mathbf{w}_k x_k$, leaving the spatial-null residual $I_k=\sum_{j\ne k}|\mathbf{g}_k^{\!\top}\mathbf{w}_j|^2$,
\begin{equation}
R_k=\log_2\!\left(1+\frac{|\mathbf{g}_k^{\!\top}\mathbf{w}_k|^2}{\sum_{j\ne k}|\mathbf{g}_k^{\!\top}\mathbf{w}_j|^2+\sigma^2}\right).
\label{eq:sdma}
\end{equation}
\item[(iv)] \emph{RSMA} splits each message into a common part (decoded by all, then canceled) and a private part, interpolating between SDMA (treat the residual as noise) and NOMA (fully decode it); user $k$ attains $R_k=C_k+R_{p,k}$ with $\sum_k C_k$ bounded by the common-stream capacity.
\end{enumerate}

Writing all four in the single form \eqref{eq:sinr} exposes its payoff: \emph{almost every} realistic impairment acts on \eqref{eq:sinr} in one of two ways:
\begin{itemize}
\item \emph{Channel perturbation ($\Delta g$):} the effective channel is corrupted, $g_k\!\to\! g_k+\Delta g_k$ (amplitude, phase, or gain \emph{ordering}). These are \emph{generic link} impairments---they exist even for a single user---so they enter the multiple-access story only through their \emph{differential} severity: a scheme suffers in proportion to how much it derives user separation from the channel itself. Power-domain access is hit hardest, because it builds both the power split and the SIC order out of $|g|$.
\item \emph{Orthogonality leakage ($\varepsilon$):} the residual rises from $I_k\!\approx\!0$ to a leakage term, $I_k\!\to\!\varepsilon$. ``Orthogonality between users'' has no meaning for a single user, so these are \emph{intrinsically multiple-access} impairments, and each corrupts the resource of a specific domain---time, frequency, space, mode, or power.
\end{itemize}
Of the impairments cataloged in Table~\ref{tab:impair}, we carry forward only the \emph{multiple-access--critical} ones: the $\varepsilon$-type impairments (one per domain) and the $\Delta g$-type impairments that are simultaneously PASS-specific and decision-changing. Impairments that degrade the single-user link exactly as in a conventional array---generic $\Delta g$ such as power-amplifier nonlinearity, oscillator phase noise, quantization, and mutual coupling---are inherited from classical analysis and noted only briefly, as is IWA \cite{Ref_Xu2025InWGAtten}, which acts on every scheme alike; those concerning hardware reliability rather than communication performance (actuator wear, impedance-mismatch reflections) are out of scope. The impairments that survive this selection are precisely the ones treated one per domain in the subsections that follow (Sections~\ref{ssec:dg}--\ref{ssec:iar}) and summarized in the impairment-versus-method map of Section~\ref{ssec:meta}; the remaining rows of Table~\ref{tab:impair} are carried no further.

\subsection{Impact of Individual Impairments}\label{ssec:perdomain}
Applying this framework, we treat the surviving impairments one at a time---the channel-perturbing pair, then one orthogonality-leakage impairment per separation domain, and finally the structural shared-guide-uplink term.

\subsubsection{Channel-Perturbing Impairments ($\Delta g$): Position Error and the Directional Pattern}\label{ssec:dg}
The dominant PASS-specific $\Delta g$ is \emph{antenna-position error}: a pinch does not sit exactly where the design places it. How the error arises depends on the activation mechanism. Under continuous activation, the servo or stepper that slides a pinch along its track stops it with a small random error of root-mean-square (rms) value $\sigma_x$, from finite resolution, drift or vibration, which the transmitter does not know. Under discrete activation nothing moves, but the desired position generally falls between two pre-installed candidates, so the nearest active pinch sits up to half a candidate spacing away. This offset is known in advance but cannot be removed, and it shrinks only as the grid becomes denser, which is why discrete activation needs dense grids. Semi-continuous activation trims this offset with a micro-actuator and so returns to a small random error. In every case, a displacement $\Delta x_n$ of pinch $n$ along the guide changes the length of both paths of its signal. The path inside the guide, $d_n$ in \eqref{eq:phase}, grows by $\Delta x_n$, and the path through the air to user $k$, $r_{nk}$, by about $\Delta x_n\cos\theta_{nk}$, where $\theta_{nk}$ is the angle between the guide, pointing away from the feed, and the line from the user to the pinch \cite{Ref_Jiang2026PositionError}. By \eqref{eq:phase}, the phase of the pinch therefore rotates by
\begin{equation}
\begin{aligned}
\Delta\phi_{nk}&\approx\frac{2\pi}{\lambda_g}\Delta x_n+\frac{2\pi}{\lambda}\Delta x_n\cos\theta_{nk}\\
&=\frac{2\pi}{\lambda}\big(n_\eff+\cos\theta_{nk}\big)\Delta x_n.
\end{aligned}
\label{eq:dphi}
\end{equation}
Because the wavelength is short, the rotation is large. The guided path alone contributes $(2\pi/\lambda)n_\eff\Delta x_n$, so even a pinch directly above its user ($\cos\theta_{nk}=0$) rotates by about $47^\circ$ per millimeter at $28$\,GHz, and by up to about $80^\circ$ when the user lies along the guide on the feed side ($\cos\theta_{nk}\to1$). This is why a $\sim$1\,mm shift can move the BER by $\sim$10\,dB \cite{Ref_AlaaEldin2026BERNOMA}.

We now analyze how the phase error enters the rate \eqref{eq:sinr}. Pinch $n$ contributes to the channel $g_k$ in \eqref{eq:sinr} the term $g_{nk}=\big(c/(4\pi f_c r_{nk}\sqrt{N})\big)e^{-j\phi_{nk}}$, with $\phi_{nk}$ from \eqref{eq:phase}. A millimeter displacement barely changes the amplitude $|g_{nk}|$, so the error acts through the phase alone. Each pinch has its own phase error $\Delta\phi_{nk}$, and the channel becomes
\begin{equation}
\tilde g_k=\sum_{n=1}^{N}g_{nk}\,e^{-j\Delta\phi_{nk}}.
\label{eq:gtilde}
\end{equation}
When the pinches are phase-aligned at user $k$, as \eqref{eq:phase} requires for coherent combining, all terms share one phase, so $|g_k|=\sum_n|g_{nk}|$. The triangle inequality then gives
\begin{equation}
|\tilde g_k|\le\sum_{n=1}^{N}\big|g_{nk}\,e^{-j\Delta\phi_{nk}}\big|=\sum_{n=1}^{N}|g_{nk}|=|g_k|,
\label{eq:triangle}
\end{equation}
with equality only if all errors $\Delta\phi_{nk}$ are equal. The signal term $|g_k|^2P_k$ of \eqref{eq:sinr} can therefore only fall, and with it the rate of every scheme that aligns several pinches at its user, such as time division re-pinched per slot. When the errors are independent and spread over the whole circle, the cross terms of $|\tilde g_k|^2$ average out. The mean power then falls to the incoherent sum $\sum_n|g_{nk}|^2$, a loss of up to a factor $N$ for $N$ equally strong pinches, as classical antenna-tolerance theory predicts \cite{Ref_Ruze1966}.
\begin{remark}[Single-pinch immunity]\label{rem:single}
For $N=1$, \eqref{eq:triangle} holds with equality, since $|\tilde g_k|=|g_{1k}e^{-j\Delta\phi_{1k}}|=|g_{1k}|=|g_k|$. A single pinch per user, as in single-pinch time division, loses nothing: position error rotates its signal but leaves $\gamma_k$ and the rate in \eqref{eq:sinr} unchanged \cite{Ref_Jiang2026PositionError}.
\end{remark}
The error also raises the interference term $I_k$ of \eqref{eq:sinr} whenever a scheme removes interference using the channel it knew before the pinches moved. In NOMA, the strong user $s$ of the pair in \eqref{eq:nomaw}--\eqref{eq:nomas} subtracts the weak user's signal with a replica built from $g_s$, whereas the received copy passed through $\tilde g_s$. The residual adds $I_s=\alpha_wP|\tilde g_s-g_s|^2$ to \eqref{eq:sinr}, so the rate \eqref{eq:nomas} becomes $\log_2(1+\tilde\gamma_s)$ with
\begin{equation}
\tilde\gamma_s=\frac{\alpha_sP|\tilde g_s|^2}{\sigma^2+\alpha_wP|\tilde g_s-g_s|^2}.
\label{eq:nomares}
\end{equation}
For one pinch, $|\tilde g_s-g_s|^2=|g_s|^2|e^{-j\Delta\phi}-1|^2=4|g_s|^2\sin^2(\Delta\phi/2)$, so a fraction $4\sin^2(\Delta\phi/2)$ of the cancelled power survives. Because the residual grows with $P$ like the signal, the SINR saturates as $P$ grows:
\begin{equation}
\lim_{P\to\infty}\tilde\gamma_s=\frac{\alpha_s|\tilde g_s|^2}{\alpha_w|\tilde g_s-g_s|^2}=\frac{\alpha_s}{4\alpha_w\sin^2(\Delta\phi/2)},
\label{eq:nomalim}
\end{equation}
where the second equality holds for one pinch. The rate therefore hits a ceiling that no increase of power removes \cite{Ref_Jiang2026PositionError}: the channel perturbation $\Delta g$ has turned into orthogonality leakage $\varepsilon$. In SDMA likewise, the nulls a precoder places with the nominal channels no longer cancel for the rotated channels, and the leakage in \eqref{eq:sdma} grows with $P$. Re-estimating the channel after the pinches land replaces $g_s$ by $\tilde g_s$ and removes this residual. It cannot undo \eqref{eq:triangle}, because a passive pinch cannot be re-phased without moving it again.

A second, recently quantified $\Delta g$ is the \emph{directional radiation pattern}. Most works model a pinch as a point that radiates equally in all directions, but coupled-mode analysis, confirmed by full-wave simulation, shows that a real pinch radiates directionally \cite{Ref_Zubair2026Pattern}, in a pencil-like beam \cite{Ref_Zhang2025DiPASS}. In \eqref{eq:effchan}, the pattern weights the term of pinch $n$ by its amplitude $a_{nk}$ in the direction of user $k$, so the signal term of \eqref{eq:sinr} holds $\tilde g_k=\sum_n a_{nk}g_{nk}$ in place of $g_k$. Placements optimized under the omnidirectional assumption, $a_{nk}=1$, therefore lose several decibels of effective gain \cite{Ref_Zubair2026Pattern}. The orientation of a pinch also becomes a design choice, and for a single pinch the best placement and orientation are known in closed form \cite{Ref_Zhang2025DiPASS}. Like position error, the pattern hits power-domain NOMA and SDMA hardest, because a design based on $g_k$ leaves the residual $\tilde g_k-g_k$ in the interference term $I_k$, as in \eqref{eq:nomares}. In NOMA, the true channel gains also differ from those the design assumed, which upsets the power split and the SIC order. In SDMA, the precoder's nulls miss the true channels. Orthogonal access is comparatively forgiving, since each user is served alone on its own resource, so an error in one user's channel does not spill into another user's signal.

\subsubsection{Reconfiguration Latency and Jitter ($\varepsilon$, time) $\to$ TDMA}
Time division with per-slot re-pinching is the scheme most exposed to the \emph{mechanical} nature of PASS: before each user is served, the pinches are moved or switched to the positions that suit this user. Unlike the electronic switching of conventional wireless systems, a re-pinch is a \emph{mechanical} operation, so it is not instantaneous: it takes a \emph{reconfiguration latency} $T_{\mathrm{rec}}$, from milliseconds to seconds for continuous (servo) actuation and from microseconds to milliseconds for discrete switching \cite{Ref_Liu2025Tutorial}. On top of this latency, the limited accuracy of the clock and of the actuator control adds a random \emph{timing jitter} $\Delta t$, so the pinches are ready slightly later or earlier than scheduled. Each re-pinch also consumes actuation energy, on top of the per-pinch activation power that energy-efficiency designs count \cite{Ref_Zhang2026EEDiscreteActivation}; it leaves the rate \eqref{eq:sinr} unchanged but adds to the consumed power, and so lowers the energy efficiency. These effects grow with the re-pinching rate, so they strike time division with per-slot re-pinching and spare the schemes that keep their pinches fixed. This matters for the field's central debate, NOMA versus orthogonal access on one waveguide. The results that make TDMA or alternating transmission optimal or superior \cite{Ref_Chen2025SumCapacityMAC,Ref_Liu2025WPTMECTDMAvsNOMA,Ref_Ren2025NOMAorOMA} all assume \emph{instantaneous, perfect} per-slot re-pinching. The first uplink design, in contrast, fixed the pinch positions over the whole frame, citing the complexity and hardware cost of per-slot re-pinching \cite{Ref_Tegos2025UplinkMinRate}.

In time division, user $k$ is served alone in its slot, so $I_k=0$ in \eqref{eq:sinr}. With $K$ equal slots per frame, its rate is $R_k=\frac{1}{K}\log_2(1+\gamma_k)$ with $\gamma_k=|g_k|^2P/\sigma^2$. While the pinches move, the transmitter must wait, since a signal sent during a re-pinch would leave through pinches still placed for the previous user. Each slot therefore begins with a guard time $T_{\mathrm{rec}}$ that carries no data. A slot lasts $T_{\mathrm{frame}}/K$, where $T_{\mathrm{frame}}$ is the frame length, so the rate becomes
\begin{equation}
\begin{aligned}
\tilde R_k&=\Big(\frac{1}{K}-\frac{T_{\mathrm{rec}}}{T_{\mathrm{frame}}}\Big)\log_2(1+\gamma_k)\\
&=\frac{1}{K}\Big(1-\frac{KT_{\mathrm{rec}}}{T_{\mathrm{frame}}}\Big)\log_2(1+\gamma_k)=\eta R_k,
\end{aligned}
\label{eq:guard}
\end{equation}
where $\eta=1-KT_{\mathrm{rec}}/T_{\mathrm{frame}}$ is the fraction of the frame left for data, a pre-log factor that the latency reduces much as pilot overhead does in training-based systems. Latency thus leaves the SINR untouched and scales the whole rate. Timing jitter adds to the latency. If the pinches are ready $\Delta t$ late, the data sent in this interval is radiated before they are in place and is lost. The guard time must therefore cover the latest arrival, so \eqref{eq:guard} holds with $T_{\mathrm{rec}}+\Delta t_{\max}$ in place of $T_{\mathrm{rec}}$, where $\Delta t_{\max}$ is the largest timing error. Data flows only if $\eta>0$, that is, only if the guard time is shorter than a slot. Keeping the lead over a scheme with fixed pinches requires more.
\begin{remark}[Latency budget]
Let $R_{\mathrm{TD}}$ be the rate of time division with instantaneous re-pinching, and $R_{\mathrm{fix}}<R_{\mathrm{TD}}$ that of a scheme with fixed pinches. By \eqref{eq:guard}, time division stays ahead, $\eta R_{\mathrm{TD}}>R_{\mathrm{fix}}$, if and only if
\begin{equation}
T_{\mathrm{rec}}<\frac{T_{\mathrm{frame}}}{K}\Big(1-\frac{R_{\mathrm{fix}}}{R_{\mathrm{TD}}}\Big).
\label{eq:trec}
\end{equation}
The guard time may thus take only a fraction of a slot, equal to the relative rate gain that re-pinching brings. With timing jitter, the budget applies to $T_{\mathrm{rec}}+\Delta t_{\max}$.
\end{remark}
For example, if re-pinching raises the rate by $10\%$, a re-pinch may take at most about $9\%$ of a slot, or $0.45$\,ms for a $10$\,ms frame shared by two users. Discrete switching, at microseconds to milliseconds, can meet such a budget; continuous actuation, at milliseconds to seconds, cannot.

Delay-sensitive traffic tightens the first condition. A real-time service serves each user once per frame, so the frame must fit within the one-way delay budget, and a slot lasts at most a $K$-th of this budget. The budget ranges from seconds for streaming down to about $1$\,ms, at or below a 5G New Radio slot, for ultra-reliable low-latency communications (URLLC) \cite{Ref_ITURM2410}, factory motion-control \cite{Ref_3GPP22104}, vehicular \cite{Ref_3GPP22186}, and haptic-teleoperation traffic \cite{Ref_Fettweis2014Tactile}. Conventional systems switch slots \emph{electronically} in nanoseconds to microseconds, whereas continuous actuation must \emph{physically move} a pinch, which is three to nine orders of magnitude slower. For such traffic, a mechanical re-pinch lasts as long as a slot or longer, so $\eta\le0$ and no time is left for data: per-slot re-pinching by mechanical movement cannot serve such traffic. Shortening the guard time does not help, since the pinches are then not yet in place when the data begins, and the start of the slot is lost. Serving such traffic therefore calls for faster reconfiguration, such as electronic switching among pre-installed pinches or waveguide segments \cite{Ref_Ouyang2025SWAN}, or for designs that avoid per-slot re-pinching altogether. Both remain open for future work.

\subsubsection{Dispersion and Beam Squint ($\varepsilon$, frequency) $\to$ FDMA/OFDMA}
Frequency-division access separates the users in frequency: plain \emph{FDMA} gives each user a disjoint sub-band, and \emph{OFDMA} gives each user a set of narrow, mutually orthogonal subcarriers. Unlike the elements of a conventional array, the pinches of a waveguide are fed in series and can sit meters apart, so the delays $\tau_{nk}=(n_\eff d_n+r_{nk})/c$ of their signals at user $k$ differ by up to nanoseconds. Even in pure line of sight, one waveguide is therefore a frequency-selective channel that delivers several delayed copies of every symbol, which cause inter-symbol interference once the band is wide \cite{Ref_Oikonomou2025OFDMA}. The delays also cause \emph{beam squint}: the point where the signals of the pinches add in phase changes with the frequency. The pinch positions place this point at the user for the carrier frequency, so at the other frequencies of the band it lies elsewhere and the user is served off the peak. To see this, write the phase in \eqref{eq:phase} as $\phi_{nk}=2\pi f_c\tau_{nk}$. At an offset $\Delta f$ from the carrier it becomes
\begin{equation}
\phi_{nk}(\Delta f)=2\pi(f_c+\Delta f)\tau_{nk}=\phi_{nk}+2\pi\Delta f\,\tau_{nk}.
\label{eq:phif}
\end{equation}
The second term, $2\pi\Delta f\,\tau_{nk}$, is the \emph{squint phase} of pinch $n$. The placement aligns the carrier phases $\phi_{nk}$, but the squint phases differ from pinch to pinch, so pinches $n$ and $n'$ arrive out of phase by $2\pi\Delta f(\tau_{nk}-\tau_{n'k})$ at the user. This squint is \emph{geometric}: it follows from the path lengths alone, even when $n_\eff$ is constant over the band. The dispersion of the guide, a change of $n_\eff$ with frequency, adds to it. The coupling coefficient $\kappa$ in \eqref{eq:cmt} depends on frequency as well, so the share $\delta_n=\sin(\kappa L_n)$ that a pinch radiates varies over the band, and with it the amplitude of its term in the signal term of \eqref{eq:sinr}. These effects grow with the bandwidth, and the squint also with the span that the pinches cover. They matter most at terahertz frequencies, where the bands are widest \cite{Ref_Jiang2026SwitchedFeed}.

We first show how the beam squint lowers the signal term of \eqref{eq:sinr}. By \eqref{eq:phif}, pinch $n$ contributes at the offset $\Delta f$ the term $\big(c/(4\pi(f_c+\Delta f)r_{nk}\sqrt{N})\big)e^{-j(\phi_{nk}+2\pi\Delta f\,\tau_{nk})}=\big(f_c/(f_c+\Delta f)\big)\,g_{nk}\,e^{-j2\pi\Delta f\,\tau_{nk}}$. For $\Delta f\ll f_c$, the factor $f_c/(f_c+\Delta f)$ is close to one and common to all pinches. Keeping the amplitudes at their carrier values, user $k$ sees at the offset $\Delta f$ the channel
\begin{equation}
g_k(\Delta f)=\sum_{n=1}^{N}g_{nk}\,e^{-j2\pi\Delta f\,\tau_{nk}}.
\label{eq:gf}
\end{equation}
At the carrier, the placement aligns the pinches, so all terms $g_{nk}$ share one phase and $|g_k|=\sum_n|g_{nk}|$. At the offset $\Delta f$, each term is rotated by its own squint phase, and
\begin{equation}
\begin{aligned}
|g_k(\Delta f)|&=\Big|\sum_{n=1}^{N}g_{nk}\,e^{-j2\pi\Delta f\,\tau_{nk}}\Big|\\
&\le\sum_{n=1}^{N}\big|g_{nk}\,e^{-j2\pi\Delta f\,\tau_{nk}}\big|=\sum_{n=1}^{N}|g_{nk}|=|g_k|,
\end{aligned}
\end{equation}
with equality only if all squint phases are equal up to multiples of $2\pi$. For a user served alone on its frequencies, $I_k=0$, and the SNR at the offset $\Delta f$ is therefore
\begin{equation}
\gamma_k(\Delta f)=\frac{|g_k(\Delta f)|^2P}{\sigma^2}\le\frac{|g_k|^2P}{\sigma^2}.
\label{eq:snrf}
\end{equation}
The SNR is thus highest at the carrier, and the squint can only lower it over the rest of the band. Two equally strong pinches, for example, have $|g_k|=2|g_{1k}|$ and $|g_k(\Delta f)|^2=|g_{1k}|^2\big|1+e^{-j2\pi\Delta f(\tau_{2k}-\tau_{1k})}\big|^2=|g_k|^2\cos^2\!\big(\pi\Delta f(\tau_{1k}-\tau_{2k})\big)$: they add fully at the carrier and cancel each other at $\Delta f=1/(2|\tau_{1k}-\tau_{2k}|)$. In general, the loss sets in once the squint phases of the pinches differ by about one radian, which defines the coherence bandwidth.
\begin{remark}[Coherence bandwidth]
The coherence bandwidth $B_c$ is the offset up to which the channel stays close to its value at the carrier. The squint phases $2\pi\Delta f\,\tau_{nk}$ have the rms spread $2\pi\Delta f\,\sigma_\tau$ over the pinches, where $\sigma_\tau$ is the rms spread of the delays. This spread stays below one radian for $\Delta f<1/(2\pi\sigma_\tau)$, which gives $B_c\approx1/(2\pi\sigma_\tau)$. For a user below the pinches, the air paths $r_{nk}$ are nearly equal, so the delays differ only through the paths $d_n$ in the guide. For many pinches spread evenly over a span $S$ along the guide, the delays are then spread evenly over an interval of width $n_\eff S/c$. Values spread evenly over an interval of width $W$ have the rms spread $W/\sqrt{12}$, so $\sigma_\tau\approx n_\eff S/(\sqrt{12}\,c)$ and
\begin{equation}
B_c\approx\frac{1}{2\pi\sigma_\tau}\approx\frac{\sqrt{12}\,c}{2\pi n_\eff S}.
\label{eq:bc}
\end{equation}
With $n_\eff=1.4$ and $S=0.6$\,m, $\sigma_\tau\approx0.81$\,ns and $B_c\approx200$\,MHz. Doubling the span halves $B_c$.
\end{remark}
We then show how the delayed copies raise the interference term of \eqref{eq:sinr}. Whether they do depends on the receiver. OFDMA keeps its subcarriers much narrower than $B_c$ and uses a cyclic prefix that covers the delay differences, so each subcarrier sees a flat channel without interference: $I_k=0$, and only the loss \eqref{eq:snrf} of the signal term remains. A single-carrier signal of bandwidth $B$ has symbols of length $1/B$. If $B$ exceeds $B_c$, the delays of the pinches differ by a large part of a symbol or more, and the receiver sees the channel \eqref{eq:gf} in time as several taps $h_{k,\ell}$, $\ell=0,\pm1,\dots$, one per symbol period. A receiver that equalizes only one tap, $\ell=0$, suffers the others as interference, $I_k=P\sum_{\ell\ne0}|h_{k,\ell}|^2$, and \eqref{eq:sinr} gives
\begin{equation}
\lim_{P\to\infty}\gamma_k=\frac{|h_{k,0}|^2}{\sum_{\ell\ne0}|h_{k,\ell}|^2}.
\label{eq:isi}
\end{equation}
Signal and interference both grow with $P$, so the rate hits a ceiling that power cannot lift, the frequency-domain form of the leakage $\varepsilon$. An ideal equalizer of all taps avoids it and leaves only the loss \eqref{eq:snrf}, as in OFDMA.

\subsubsection{Crosstalk and Synchronization Errors ($\varepsilon$, space and mode) $\to$ SDMA and MDM}
Space division across waveguides feeds each guide from its own RF chain. Waveguide multiplexing precodes the streams across the guides and places spatial nulls, so that each user receives only its own stream \cite{Ref_Zhao2025TransmissionStructures,Ref_Bereyhi2025MIMO,Ref_MIMOPASS2025DLUL}. The nulls hold only if the channels of all guides are known and the RF chains are aligned in phase and time. Crosstalk and misalignment violate this. \emph{Crosstalk}, the inter-waveguide coupling, means that part of the signal in one guide leaks into a neighboring guide. \emph{Misalignment} is a synchronization error: the phases and the timing of the RF chains drift apart between calibrations.

In \eqref{eq:multiwg}, guide $m$ carries the signal $s_m$ that the precoder assigns to it. In practice, two errors act on this signal before it reaches the pinches. First, the RF chain of guide $m$ adds a phase error $\theta_m$, so the guide is actually fed with $s_me^{j\theta_m}$. A timing error of the RF chain has the same effect and is included in $\theta_m$, since a delay much shorter than a symbol rotates the carrier by $2\pi f_c$ times the delay. Second, crosstalk passes a part of the signal of each guide to the other guides. The signal fed into guide $m$ travels in guide $m'$ with the complex amplitude $T_{m',m}$, which we take as equal for all pinches of guide $m'$. Let $\mathbf{D}=\mathrm{diag}(e^{j\theta_1},\dots,e^{j\theta_M})$ collect the phase errors, and let the coupling matrix $\mathbf{T}$ collect the amplitudes $T_{m',m}$, with $\mathbf{T}=\mathbf{I}$ without crosstalk. Then, \eqref{eq:multiwg} becomes
\begin{equation}
y_k=\tilde{\mathbf{g}}_k^{\!\top}\mathbf{s}+w_k,\qquad \tilde{\mathbf{g}}_k^{\!\top}=\mathbf{g}_k^{\!\top}\mathbf{T}\mathbf{D}.
\label{eq:xtalk}
\end{equation}
User $k$ sees the true channel $\tilde{\mathbf{g}}_k$, whereas the precoder is designed for the nominal channel $\mathbf{g}_k$, since the transmitter does not know the errors. Let the precoder be zero-forcing, $\mathbf{g}_k^{\!\top}\mathbf{w}_j=0$ for $j\ne k$, which needs $M\ge K$ guides, and let the $K$ streams share the power $P$ equally, $\mathbf{w}_j=\sqrt{P/K}\,\mathbf{v}_j$ with unit-norm $\mathbf{v}_j$. With the channel error $\Delta\mathbf{g}_k=\tilde{\mathbf{g}}_k-\mathbf{g}_k$, stream $j\ne k$ reaches user $k$ with the amplitude
\begin{equation}
\tilde{\mathbf{g}}_k^{\!\top}\mathbf{w}_j=\mathbf{g}_k^{\!\top}\mathbf{w}_j+\Delta\mathbf{g}_k^{\!\top}\mathbf{w}_j=\sqrt{P/K}\,\Delta\mathbf{g}_k^{\!\top}\mathbf{v}_j,
\label{eq:leak}
\end{equation}
since the nulls cancel the nominal part only. Therefore, the SINR of user $k$ in \eqref{eq:sdma} is rewritten as
\begin{equation}
\tilde\gamma_k=\frac{(P/K)\,|\tilde{\mathbf{g}}_k^{\!\top}\mathbf{v}_k|^2}{\sigma^2+(P/K)\sum_{j\ne k}|\Delta\mathbf{g}_k^{\!\top}\mathbf{v}_j|^2}.
\label{eq:zfsinr}
\end{equation}
Signal and interference both grow with $P$, so the rate hits a ceiling that power cannot lift, as known for multi-antenna links with transceiver impairments \cite{Ref_Bjornson2013Impairments}. The ceiling follows from \eqref{eq:zfsinr} as
\begin{equation}
\lim_{P\to\infty}\tilde\gamma_k=\frac{|\tilde{\mathbf{g}}_k^{\!\top}\mathbf{v}_k|^2}{\sum_{j\ne k}|\Delta\mathbf{g}_k^{\!\top}\mathbf{v}_j|^2}.
\label{eq:zflim}
\end{equation}
It tends to fall as more streams are served, since each adds a term to the sum.
\begin{remark}[Waveguide division]\label{rem:wd}
Waveguide division feeds stream $j$ into guide $j$ alone and uses no precoder across the guides, with user $k$ served by guide $k$. Substituting $\mathbf{w}_j=\sqrt{P/K}\,\mathbf{e}_j$, with the $j$-th unit vector $\mathbf{e}_j$, into \eqref{eq:leak}, we have
\begin{equation}
\tilde{\mathbf{g}}_k^{\!\top}\mathbf{w}_j=\sqrt{P/K}\,\tilde{\mathbf{g}}_k^{\!\top}\mathbf{e}_j=\sqrt{P/K}\,\tilde g_{k,j},
\end{equation}
where $\tilde g_{k,j}=\big(\sum_{m}g_{k,m}T_{m,j}\big)e^{j\theta_j}$ is the $j$-th entry of the true channel \eqref{eq:xtalk}. Unlike in \eqref{eq:leak}, no nulls cancel its nominal part. Then, the SINR of user $k$ in \eqref{eq:sdma} is rewritten as
\begin{equation}
\tilde\gamma_k=\frac{(P/K)\,|\tilde g_{k,k}|^2}{\sigma^2+(P/K)\sum_{j\ne k}|\tilde g_{k,j}|^2}.
\label{eq:wd}
\end{equation}
The phase errors cancel in $|\tilde g_{k,j}|$, so waveguide division is immune to a misalignment of the RF chains. Crosstalk can harm it, since $\tilde g_{k,j}$ contains the term $g_{k,k}T_{k,j}$: a share of stream $j$ arrives through the user's own guide, and no precoder cancels it. Even without errors, where $\tilde g_{k,j}=g_{k,j}$, \eqref{eq:wd} saturates at $|g_{k,k}|^2/\sum_{j\ne k}|g_{k,j}|^2$ as $P$ grows. This ceiling is set by the leakage through the air, which the placement of the pinches can keep small \cite{Ref_Ding2025WDMA}.
\end{remark}

\emph{Mode division} multiplexes the streams on different guided modes of one waveguide, with one RF chain per mode and each pinch matched to the mode it should radiate \cite{Ref_Xu2026MultiModeMDM}. The modes are orthogonal only in an ideal guide. A matched pinch also radiates some power of the other modes \cite{Ref_Xu2026MultiModeMDM}, and bends of the guide or a polarization mismatch couple power between them. The model \eqref{eq:xtalk} then holds with $\mathbf{T}$ describing the coupling between the modes, and the leakage caps the rate as in \eqref{eq:zflim}.

\begin{table*}[!t]
\renewcommand{\arraystretch}{1.3}
\caption{Impairment-Versus-Method Severity Map: $\bullet\bullet\bullet$ critical, $\bullet\bullet$ moderate, $\bullet$ minor, -- not applicable. Rows: channel perturbations ($\Delta g$), orthogonality leakages ($\varepsilon$), and inter-antenna radiation.}
\label{tab:matrix}
\centering
\footnotesize
\begin{tabular}{@{}p{4.6cm}ccccccc@{}}
\toprule
\rowcolor{headblue}\textcolor{white}{\textbf{Impairment (type $\cdot$ domain)}} & \textcolor{white}{\textbf{TDMA}} & \textcolor{white}{\textbf{OFDMA}} & \textcolor{white}{\textbf{NOMA}} & \textcolor{white}{\textbf{RSMA}} & \textcolor{white}{\textbf{SDMA}} & \textcolor{white}{\textbf{WD/EDMA}} & \textcolor{white}{\textbf{MDM}} \\
\midrule
Position error / directional pattern ($\Delta g$) & $\bullet\bullet$ & $\bullet\bullet$ & $\bullet\bullet\bullet$ & $\bullet\bullet$ & $\bullet\bullet\bullet$ & $\bullet\bullet$ & $\bullet\bullet$ \\
\midrule
Reconfig.\ latency / jitter ($\varepsilon\cdot$time) & $\bullet\bullet\bullet$ & $\bullet$ & $\bullet$ & $\bullet$ & $\bullet$ & $\bullet$ & $\bullet$ \\
Dispersion / beam squint ($\varepsilon\cdot$freq.) & $\bullet$ & $\bullet\bullet$ & $\bullet\bullet$ & $\bullet\bullet$ & $\bullet\bullet$ & $\bullet\bullet$ & $\bullet\bullet$ \\
Inter-WG crosstalk / sync ($\varepsilon\cdot$space) & -- & -- & $\bullet\bullet$ & $\bullet\bullet$ & $\bullet\bullet\bullet$ & $\bullet$ & $\bullet\bullet$ \\
Mode / polarization crosstalk ($\varepsilon\cdot$mode) & -- & -- & -- & -- & $\bullet$ & -- & $\bullet\bullet\bullet$ \\
Imperfect SIC / order error ($\varepsilon\cdot$power) & -- & -- & $\bullet\bullet\bullet$ & $\bullet\bullet$ & $\bullet\bullet$ & $\bullet\bullet$ & -- \\
\midrule
Inter-antenna radiation (UL) & $\bullet\bullet$ & $\bullet\bullet$ & $\bullet\bullet$ & $\bullet\bullet$ & $\bullet\bullet$ & $\bullet\bullet$ & $\bullet$ \\
\bottomrule
\end{tabular}
\end{table*}
\subsubsection{Imperfect SIC and SIC-Order Errors ($\varepsilon$, power) $\to$ NOMA}
Power-domain NOMA relies on SIC, so imperfect cancellation is its native impairment. The replica that the strong user subtracts may not match the received signal, because the channel knowledge is inaccurate or outdated, or because the weak user's message was decoded in error. A share of the cancelled signal then remains and produces an error floor at high SNR \cite{Ref_Yue2026NOMAPerformance,Ref_AlaaEldin2026BERNOMA}. The cancellation also fails when the SIC order is wrong, and in PASS an impairment of the pinches can invert the order, which is read from the channel gains.

SIC works with the channel $g_s$ that is known, whereas the signal arrives through the true channel $\tilde g_s$. The two differ for several reasons. The estimate may be noisy \cite{Ref_Ye2026FairnessNOMACSI}. The channel may have changed since the estimate, because the user moved or the pinches were re-placed or displaced. The channel may also be computed from a model that ignores the radiation pattern of the pinches or, on the uplink, IAR.

The transmitter superposes the unit-power symbols $x_w$ and $x_s$ of the pair in \eqref{eq:nomaw}--\eqref{eq:nomas}, and the strong user receives them through its true channel $\tilde g_s$,
\begin{equation}
y_s=\tilde g_s\big(\sqrt{\alpha_wP}\,x_w+\sqrt{\alpha_sP}\,x_s\big)+w_s,
\end{equation}
with the noise $w_s$ of power $\sigma^2$. It first decodes $x_w$, rebuilds the weak user's signal with the channel $g_s$ that it knows, and subtracts this replica:
\begin{equation}
\begin{aligned}
y_s-g_s\sqrt{\alpha_wP}\,x_w&=\tilde g_s\sqrt{\alpha_sP}\,x_s\\
&\quad+(\tilde g_s-g_s)\sqrt{\alpha_wP}\,x_w+w_s.
\end{aligned}
\end{equation}
The first term is the wanted signal. The second is the residual of the cancellation, which vanishes only if the known channel equals the true one, $g_s=\tilde g_s$. Its power, $\alpha_wP|\tilde g_s-g_s|^2$, is the interference term of \eqref{eq:sinr} that gives the SINR \eqref{eq:nomares} and the ceiling \eqref{eq:nomalim}. Both therefore hold for any mismatch between the known and the true channel, and a decoding error of $x_w$ leaves a residual in the same way \cite{Ref_AlaaEldin2026BERNOMA}. The ceiling \eqref{eq:nomalim} falls as the mismatch $|\tilde g_s-g_s|$ grows. If the mismatch exceeds the channel itself, $|\tilde g_s-g_s|>|\tilde g_s|$, the subtraction adds power instead of removing it.

A wrong order makes the cancellation fail as well. The weak user's message is sent at the rate \eqref{eq:nomaw}. The strong user can decode it only if its own SINR for this message is at least that of the weak user,
\begin{equation}
\frac{\alpha_w|\tilde g_s|^2P}{\alpha_s|\tilde g_s|^2P+\sigma^2}\ge\frac{\alpha_w|g_w|^2P}{\alpha_s|g_w|^2P+\sigma^2}.
\label{eq:sicorder}
\end{equation}
The left side grows with $|\tilde g_s|^2$, so \eqref{eq:sicorder} holds if and only if $|\tilde g_s|^2\ge|g_w|^2$. If an impairment lowers the true gain of the strong user below that of the weak user, the strong user cannot decode the weak user's message, and its own message is lost as well. The order depends on the gains only, so a mismatch of the phase alone leaves it intact.

Choosing the order is an open design problem in PASS. It is trivial for two users but combinatorial in general \cite{Ref_Ai2026NOMAMECDelay}. A low-complexity dynamic ordering replaces the exhaustive search \cite{Ref_Vashakidze2025RadiationPower}, and a conservative SINR evaluation keeps SIC feasible under channel uncertainty \cite{Ref_Ye2026FairnessNOMACSI}. With several waveguides, the optimal order follows the ratio of interference plus noise to channel gain and holds under any power allocation \cite{Ref_Wang2025MultiWGActNOMA}. Rate splitting performs only one cancellation, that of the common stream, so it has fewer steps that a residual or a wrong order can affect. Schemes that run NOMA within each of several waveguides inherit both problems \cite{Ref_Wang2025MultiWGActNOMA}, whereas pure precoding across the guides performs no cancellation.

\subsubsection{Inter-Antenna Radiation $\to$ Shared-Waveguide Uplink}\label{ssec:iar}
IAR affects every uplink scheme that keeps several pinches active on one guide at once. By \eqref{eq:ul}, the wave of user $k$, captured by pinch $n'$, loses the share $\delta_i^2$ of its power at every pinch $i$ that it passes on the way to the feed, so the signal term $|g_k^{\mathrm{ul}}|^2P_k$ of \eqref{eq:sinr} holds the factor
\begin{equation}
\prod_{i<n'}\big(1-\delta_i^2\big)=\big(1-\delta^2\big)^{n'-1},
\label{eq:iargain}
\end{equation}
where the equality holds for identical pinches, $\delta_i=\delta$. The power thus falls geometrically with the number of pinches passed. In NOMA, the powers of the other users, which form the interference term, carry such factors as well.

The factor differs from user to user, since it depends on where the serving pinch sits along the guide. It therefore changes the order of the powers that arrive at the feed, on which the power split and the SIC order of NOMA rely.
\begin{remark}[IAR-free operation]\label{rem:iarfree}
If only one pinch is active per slot or per segment, the product in \eqref{eq:iargain} is empty and IAR vanishes. Time division with one pinch per slot and segmented waveguides \cite{Ref_Ouyang2025SWAN,Ref_Gu2025SWANUplink} operate in this way, the latter at the cost of one feed per segment, each with its own wired link to the base station \cite{Ref_Gu2025SWANUplink}.
\end{remark}
The factor \eqref{eq:iargain} covers the first order only. Beyond it, the pinches of a guide also interact through the waves that they radiate, and a complete end-to-end model of the uplink remains open for future work \cite{Ref_Wang2025Multiport}.

\subsection{Summary of Impacts and Severity Map}\label{ssec:meta}
The impairments enter the rate in different ways. Position error, the radiation pattern, beam squint and IAR lower the signal term of \eqref{eq:sinr}. Latency and timing jitter take away time, the factor $\eta$ in \eqref{eq:guard}; with the guard time that \eqref{eq:guard} assumes, they leave the SINR untouched. A cancellation residual, delayed copies, and crosstalk or misalignment raise the interference term. This is the most harmful way, since the rate then hits a ceiling that power cannot lift. Which way an impairment takes depends on the access scheme, and this yields the central finding.
\begin{quote}
\emph{An impairment generally harms an access scheme most when it corrupts the domain in which the scheme separates its users.}
\end{quote}
The impairment then breaks the mechanism that keeps the users apart: the time that per-slot re-pinching needs, the spatial nulls, the orthogonal modes, or the cancellation. A scheme that does not use this mechanism is affected less or not at all.

Table~\ref{tab:matrix} collects these results in a \emph{qualitative} severity map. For the leakages in time, space, mode, and power, the critical mark lies on the scheme of its own domain, the diagonal of the map. Beam squint is the exception: the cyclic prefix keeps the subcarriers of OFDMA flat, and the loss follows the spread of the pinches. A cell marked ``--'' is a scheme that the impairment cannot reach. The channel perturbations ($\Delta g$) are the only row that is at least moderate for every scheme, since every scheme needs an accurate channel. They are critical for NOMA and SDMA, where they also raise the interference term. The rate-splitting column is milder than the NOMA column, because rate splitting performs only one cancellation \cite{Ref_RSMA2025MultiWG}. The rows of imperfect SIC and of IAR rate the variants that are exposed: the space-domain schemes that run NOMA within each guide, and the uplink schemes that keep several pinches active at once. Under position error, orthogonal access with a single pinch per user is the most robust, since its rate stays unchanged.

\section{Unified Impairment-Aware Benchmark}\label{sec:bench}
This section measures the impact of the impairments with a unified benchmark that compares time-, frequency-, power\nobreakdash-, and space-domain access under identical channel and impairment models. The parameters, listed in Table~\ref{tab:simparams}, mainly follow \cite{Ref_Ding2025PinchingPerspective}. Each impairment is switched on alone, at a level stated with each result, and all schemes see the same user placements and the same random draws of the impairment. The benchmark covers the separation domains for which a common channel model exists: time, frequency, power, and space on the downlink, and inter-antenna radiation on the uplink to first order. Mode division, environment division, segmented waveguides, and multi-cell operation need further models and are not simulated.

\subsection{Benchmark Setup}
The users are placed uniformly in a square of side $D=10$\,m centered at the origin, and the guide runs parallel to the $x$-axis at height $d=3$\,m, fed at the left edge of the square. The active pinches share the power equally. For the space domain, two guides run at $y=\pm D/3$. All channels are the line-of-sight channels of \eqref{eq:effchan}, and the transmitter knows them apart from the impairment under study. Every experiment serves two users, who do not move during a frame. This is the smallest case in which users must be separated and the time shares and power splits can be optimized exactly. It is also the setting of the PASS comparisons of NOMA and OMA in \cite{Ref_Ren2025NOMAorOMA,Ref_Yue2026NOMAPerformance}. The best fixed placement is searched over two kinds of layout: all pinches in phase at one user, or one group of pinches in phase at each user. On one guide, the sum rate is largest when only the stronger user is served. We therefore measure fairness by the minimum user rate, the rate of the worse-served user. For each user placement, each scheme chooses its time shares, subcarriers, or power split to maximize this rate within the budget $P$. Each user's rate is averaged over the impairment draws before the minimum is taken, and this minimum is then averaged over the user placements. For position error, the sum rate with at least $1$\,bit/s/Hz per user is also given.

The in-waveguide attenuation is neglected, as in \eqref{eq:cmt}. For low-loss guides and areas of this size, \cite{Ref_Xu2025InWGAtten} shows that it costs a negligible rate. At $0.08$\,dB/m, the $10$\,m guide loses at most $0.8$\,dB. On one guide and without impairments, a run with this loss changes the minimum user rates by less than $0.1$\,bit/s/Hz and leaves the ranking of the schemes unchanged. Lossier guides, with up to $1.3$\,dB/m \cite{Ref_Zhang2025DiPASS}, lie outside this benchmark. Without impairments, one pinch above a single user reaches $8.8$ to $15.5$\,bit/s/Hz between $10$ and $30$\,dBm, which matches \cite{Ref_Ding2025PinchingPerspective}.

\begin{table}[!t]
\renewcommand{\arraystretch}{1.2}
\caption{Simulation Parameters.}
\label{tab:simparams}
\centering
\footnotesize
\rowcolors{2}{acrzeb}{white}
\begin{tabular}{@{}p{3.55cm}p{4.65cm}@{}}
\toprule
\rowcolor{headblue}\textcolor{white}{\textbf{Parameter}} & \textcolor{white}{\textbf{Value}} \\
\midrule
Carrier frequency $f_c$ & $28$\,GHz ($\lambda=10.71$\,mm) \\
Effective index $n_\eff$ & $1.4$ ($\lambda_g=7.65$\,mm) \\
Reference gain $c^2/(4\pi f_c)^2$ & $-61.4$\,dB \\
Noise power $\sigma^2$ & $-90$\,dBm \\
Waveguide height $d$ & $3$\,m \\
Square service area & side $D=10$\,m, centered at the origin \\
Feed point & left edge of the area, $(-D/2,0,d)$ \\
Two-guide layout & guides at $y=\pm D/3$, one pinch each \\
Transmit power $P$ & $0$--$40$\,dBm ($20$\,dBm default) \\
Power over the active pinches & equal \\
Minimum pinch spacing & $\lambda/2$ \\
In-waveguide attenuation & neglected ($0.08$\,dB/m in one check run) \\
Users $K$ & $2$, uniform in the area, static in a frame \\
Active pinches $N$ & $1$--$8$ (baseline), $2$ otherwise \\
Frame length $T_{\mathrm{frame}}$ & $10$\,ms \\
User placements & $2000$ ($10^4$ per class with two guides) \\
Impairment draws & $200$ per placement \\
\bottomrule
\end{tabular}
\end{table}

\subsection{Baseline Without Impairments}
\begin{figure}[!t]
\centering
\includegraphics[width=\columnwidth]{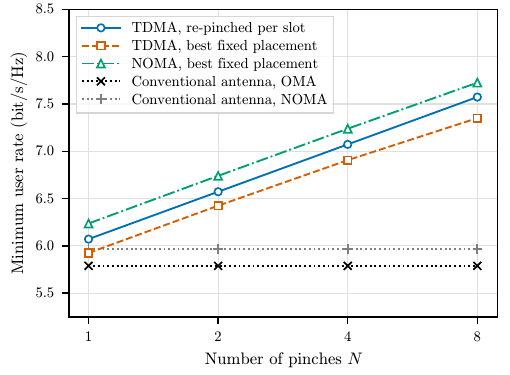}
\caption{Minimum user rate versus the number of pinches without impairments, at $P=20$\,dBm. Solid line: from \eqref{eq:arraygain}; markers: simulation.}
\label{fig:base}
\end{figure}
Fig.~\ref{fig:base} compares the schemes without impairments at $P=20$\,dBm as the number of pinches $N$ grows. With per-slot re-pinching, time division gathers all $N$ pinches above the served user, where they add in phase. By \eqref{eq:arraygain}, each doubling of $N$ then doubles the SNR of this user and adds $1$\,bit/s/Hz to its rate in its own slot. Since each user holds the guide for about half of the frame, the minimum user rate grows by $0.5$\,bit/s/Hz per doubling, as the solid line shows. Up to $N=8$, the pinches of a slot stay within about $5$\,cm of the user, so their in-phase sum outweighs the power dilution of Remark~\ref{rem:dilution}. Fixed pinches, in contrast, serve both users, and each user receives part of its signal over a longer path. With its best fixed placement, time division therefore falls $0.15$ to $0.22$\,bit/s/Hz below per-slot re-pinching.

NOMA with its best fixed placement leads per-slot re-pinching by $0.15$ to $0.17$\,bit/s/Hz, in $96$\,\% of the user placements at $N=2$. By \eqref{eq:nomaw} and \eqref{eq:nomas}, the equal-rate split at high SNR gives the strong user a share of about $(|g_s|^2P/\sigma^2)^{-1/2}$, so both users reach about half the rate that the strong user would get alone. The weak user is then limited by the interference of this share rather than by noise, and the best placement gathers all pinches above one user, mostly the one closer to the guide. Time division reaches only a rate between half the weak user's rate and half the strong user's. The lead of NOMA is therefore about a quarter of the difference between the rates that the two users would get alone, and it vanishes for users at equal distance from the guide. It also needs a high SNR. Below a minimum user rate of about $2$ to $3$\,bit/s/Hz, the weak user is limited by noise, and per-slot re-pinching is ahead. This is the regime of \cite{Ref_Ren2025NOMAorOMA}, where time division needs less power than NOMA at rate targets of $3$\,bit/s/Hz for $N\ge4$. A conventional antenna at the center of the area, $(0,0,d)$, radiating the full power as in \cite{Ref_Ding2025PinchingPerspective}, stays $0.28$\,bit/s/Hz below one re-pinched pinch with OMA and $0.27$\,bit/s/Hz below one fixed pinch with NOMA, because the pinch can move closer to the users. At $10$ and $30$\,dBm, every curve shifts by $1.66$ to $1.70$\,bit/s/Hz per $10$\,dB, and the order of the schemes is the same. All differences quoted are paired, with $95\,\%$ confidence intervals below $\pm0.01$\,bit/s/Hz.

\subsection{Position Error}
\begin{figure*}[!t]
\centering
\includegraphics[width=\textwidth]{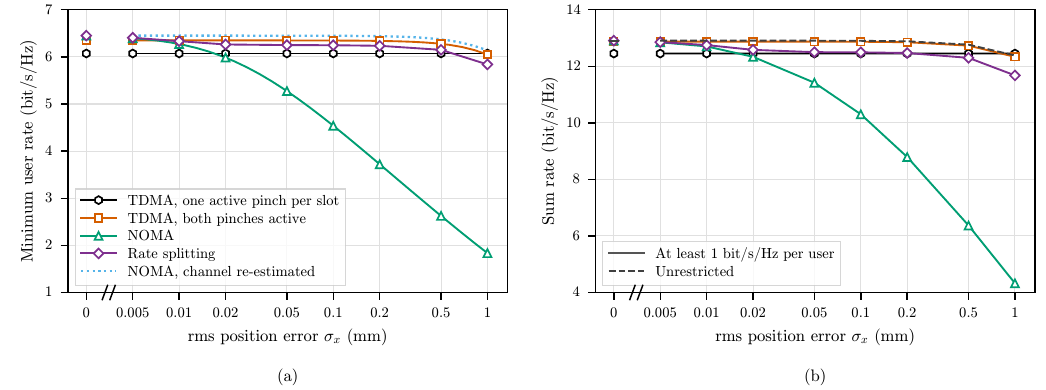}
\caption{Rates versus the rms position error at $P=20$\,dBm, one pinch above each user. (a)~Minimum user rate. (b)~Sum rate with at least $1$\,bit/s/Hz per user (solid) and unrestricted (dashed). Lines in (a), except rate splitting: from \eqref{eq:dphi}--\eqref{eq:nomares}; markers: simulation; left edge: no error.}
\label{fig:pos}
\end{figure*}
Fig.~\ref{fig:pos} shows the effect of position error at $P=20$\,dBm. The experiment uses two pinches, one above each user, which are fine-positioned within $\pm5$\,cm to maximize the smaller of the two user gains. Without error, TDMA with both pinches active then reaches $6.35$\,bit/s/Hz and NOMA $6.45$\,bit/s/Hz. Each pinch is displaced along the guide by an independent Gaussian error of rms value $\sigma_x$. The schemes choose their time shares and power splits knowing $\sigma_x$ but not the errors, and NOMA and rate splitting cancel with the channel known before the pinches moved, as in \eqref{eq:nomares}. In each draw, each message is sent at a rate that every user that must decode it supports. The strong user therefore decodes the weak user's message even where \eqref{eq:sicorder} fails. TDMA with one active pinch per slot then keeps its rate at every error, as Remark~\ref{rem:single} states. With both pinches active, the two terms of \eqref{eq:gtilde} drift apart in phase, and the rate falls as \eqref{eq:triangle} allows, by less than $0.2$\,\% up to $0.2$\,mm and by $5$\,\% at $1$\,mm.

NOMA loses far more, because the residual of the cancellation enters the interference term of the strong user, \eqref{eq:nomares}. A phase error of about one degree, reached at $0.02$\,mm, already leaves a residual above the noise. NOMA therefore falls below TDMA with both pinches beyond about $0.007$\,mm and below one active pinch beyond about $0.017$\,mm. More power helps little, as the ceiling \eqref{eq:nomalim} predicts \cite{Ref_Jiang2026PositionError}: at $0.1$\,mm, raising $P$ from $20$ to $30$\,dBm lifts the minimum user rate of NOMA by only $0.14$\,bit/s/Hz, against $1.66$\,bit/s/Hz without error. Rate splitting avoids most of this loss by moving both messages into the common stream, which each user decodes before any cancellation. It stays within about $0.13$\,bit/s/Hz of TDMA with both pinches up to $0.5$\,mm. Beyond about $0.01$\,mm, it lies below TDMA, because the common stream runs at the rate that the weaker user supports. Re-estimating the channel after the pinches land removes the residual, and NOMA then keeps a lead of about $0.10$\,bit/s/Hz up to $0.2$\,mm. The advantage of orthogonal access thus rests on a cancellation with inaccurate channel information. In panel (b), without a condition on the users, every scheme with both pinches serves the stronger user alone and reaches the same sum rate. With at least $1$\,bit/s/Hz for each user, the order of panel (a) returns: at $0.2$\,mm, NOMA reaches $8.8$\,bit/s/Hz, rate splitting $12.5$\,bit/s/Hz, and TDMA with both pinches $12.9$\,bit/s/Hz. A power-domain scheme that may skip the cancellation never does so here and coincides with NOMA.

\subsection{Reconfiguration Latency}
\begin{figure}[!t]
\centering
\includegraphics[width=\columnwidth]{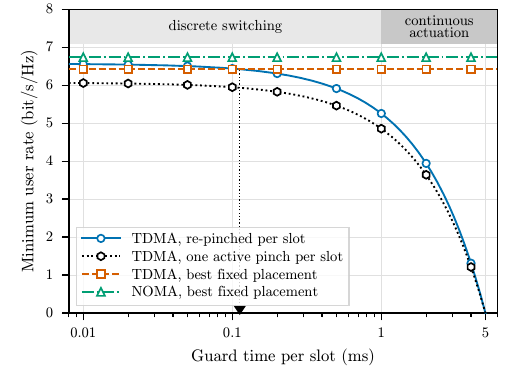}
\caption{Minimum user rate versus the guard time per slot at $P=20$\,dBm and $T_{\mathrm{frame}}=10$\,ms. Lines: from \eqref{eq:guard}; markers: simulation; tick: budget from \eqref{eq:trec}; shaded: latency ranges \cite{Ref_Liu2025Tutorial}.}
\label{fig:lat}
\end{figure}
Fig.~\ref{fig:lat} shows the minimum user rate against the guard time per slot, $T_{\mathrm{rec}}$, or $T_{\mathrm{rec}}+\Delta t_{\max}$ with timing jitter, for a $10$\,ms frame shared by two users. By \eqref{eq:guard}, the guard time scales the rate of every scheme that reconfigures per slot by the factor $\eta$ and leaves its SINR unchanged. The markers, which re-optimize the time shares for each guard time, confirm this, so the curves follow from the rates without guard time of Figs.~\ref{fig:base} and~\ref{fig:pos}. TDMA with one active pinch per slot starts below both fixed placements, since each user is served by one pinch; its merit is the immunity to position error of Fig.~\ref{fig:pos}.

Without guard time, per-slot re-pinching with two pinches leads TDMA with the best fixed placement by $2.3$\,\%. By \eqref{eq:trec}, a guard time of $0.11$\,ms consumes this lead, and the budget falls from $0.15$\,ms at $10$\,dBm to $0.09$\,ms at $30$\,dBm as the relative lead shrinks. It lies within the range of discrete switching and below that of continuous actuation \cite{Ref_Liu2025Tutorial}. A longer frame widens the budget in proportion, to $11$\,ms for a $1$\,s frame, which is still shorter than one mechanical move of $20$--$30$\,ms \cite{Ref_Bozkurt2026Trajectory}. NOMA with its best fixed placement stays ahead of per-slot re-pinching at every guard time.

\subsection{Beam Squint}
\begin{figure*}[!t]
\centering
\includegraphics[width=\textwidth]{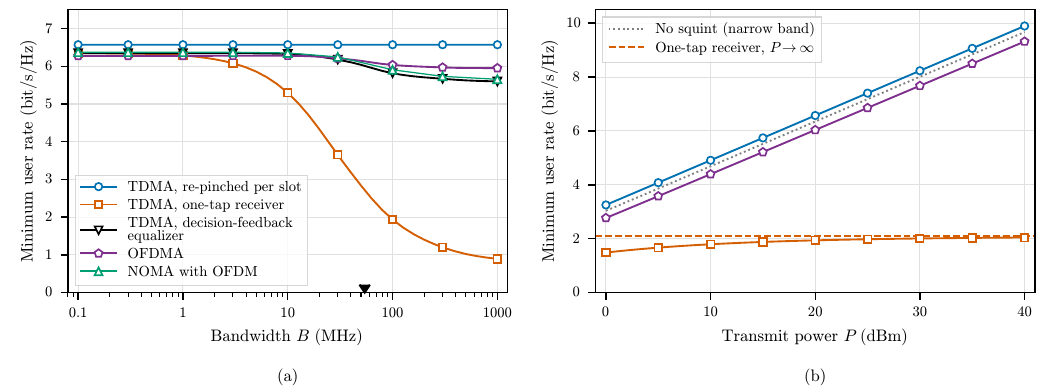}
\caption{Minimum user rate versus (a) the bandwidth at $P=20$\,dBm and (b) the transmit power at $B=100$\,MHz. Lines: from \eqref{eq:gf}--\eqref{eq:isi}, except NOMA with OFDM; markers: simulation; tick: $B=2B_c$ from \eqref{eq:bc}.}
\label{fig:bw}
\end{figure*}
Fig.~\ref{fig:bw} shows the effect of the bandwidth $B$ on the layout of Fig.~\ref{fig:pos}, with both pinches active. The noise is held at $-90$\,dBm at every bandwidth, so that only the squint acts. At a user, the delays of the two pinches differ by $11.8$\,ns in the median, so $\sigma_\tau\approx5.9$\,ns. By \eqref{eq:bc}, $B_c\approx27$\,MHz. The tick marks $B=2B_c$, where the band edge reaches $B_c$. A single-carrier receiver that equalizes one tap suffers the copy from the other pinch as interference. It uses sinc pulses and samples at the instant that maximizes its SINR. At this SNR, a delay of a small fraction of a symbol already matters: the minimum user rate loses $10$\,\% at about $6$\,MHz, halves at about $40$\,MHz and falls to $0.89$\,bit/s/Hz at $1$\,GHz. More power helps little, since at $100$\,MHz the rate approaches the ceiling \eqref{eq:isi} of $2.10$\,bit/s/Hz, as panel~(b) shows.

An ideal decision-feedback equalizer avoids this ceiling and loses only the band average of \eqref{eq:snrf}, $12$\,\% at $1$\,GHz. OFDMA uses a cyclic prefix of $100$\,ns, which covers every delay difference in the area, and a subcarrier spacing of $120$\,kHz. It loses $5$\,\% at $1$\,GHz, because it gives each user the subcarriers on which its channel is strong. At $100$\,MHz, it runs parallel to the rate without squint as the power grows. NOMA with orthogonal frequency-division multiplexing (OFDM) serves both users on every subcarrier, with one power split for the whole band and the decoding order set by the band-average gains. It loses $11$\,\% at $1$\,GHz, about as much as the decision-feedback equalizer. TDMA with per-slot re-pinching keeps its rate up to $1$\,GHz, because its two pinches sit within about one guided wavelength of each other. These results cover the geometric squint only: $n_\eff$ is constant over the band, and the coupling does not depend on frequency.

\subsection{Crosstalk and Misalignment}\label{ssec:e5}
\begin{figure*}[!t]
\centering
\includegraphics[width=\textwidth]{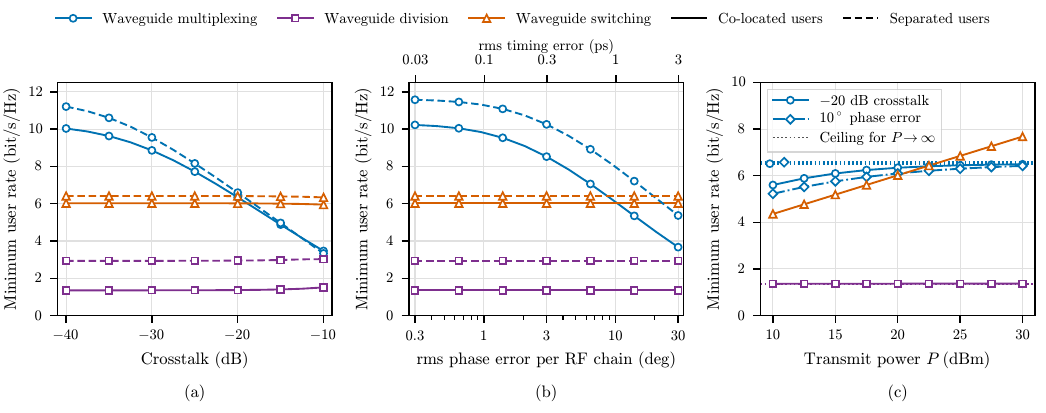}
\caption{Minimum user rate with two guides versus (a) the crosstalk and (b) the rms phase error per RF chain at $P=20$\,dBm, and (c) the transmit power for co-located users. Dotted: ceilings from \eqref{eq:zflim} and \eqref{eq:wd} for $P\to\infty$, with the optimized power split instead of equal powers.}
\label{fig:space}
\end{figure*}
Fig.~\ref{fig:space} shows the effect of crosstalk and misalignment with two guides at $y=\pm D/3$, each with one pinch at the position of the user closer to it, whom it serves in waveguide division. Co-located users lie in a central box of $4\times3$\,m, and separated users on the side of their own guide, $1.5\le|y|\le5$\,m. Crosstalk follows a lossless coupler with an unknown, uniformly distributed phase, and misalignment an independent Gaussian phase error per RF chain. Both are residuals after calibration, since a known coupling could be precoded away, and no measured values exist for PASS. Each scheme chooses its power split or time shares for the given error level. Without errors, waveguide multiplexing reaches $10.3$\,bit/s/Hz for co-located and $11.6$\,bit/s/Hz for separated users, far above switching, with $6.0$ and $6.4$\,bit/s/Hz, and division, with $1.4$ and $2.9$\,bit/s/Hz. Division is limited by the leakage through the air, its ceiling in Remark~\ref{rem:wd}.

Crosstalk fills the nulls of \eqref{eq:leak}: at $-20$\,dB, multiplexing falls to $6.3$\,bit/s/Hz for co-located and $6.6$\,bit/s/Hz for separated users, and beyond about $-19$\,dB switching is better for both. A phase error of $10^\circ$ per RF chain, which corresponds to a timing error of $1$\,ps, costs about as much as this crosstalk for co-located users, with $6.1$\,bit/s/Hz left, but less for separated users, with $8.0$\,bit/s/Hz. Division changes little with crosstalk and, as Remark~\ref{rem:wd} states, not at all with misalignment. With either error, multiplexing stops at the ceiling \eqref{eq:zflim} as the power grows, and switching overtakes it between $21$ and $22$\,dBm for co-located users. The rule of \cite{Ref_Zhao2025TransmissionStructures} that full precoding pays off most for co-located users holds here against division. Without errors, multiplexing gains $8.9$\,bit/s/Hz over division for co-located and $8.7$\,bit/s/Hz for separated users, and $5.0$ and $3.6$\,bit/s/Hz at $-20$\,dB crosstalk. At $10^\circ$, the order reverses, with $4.7$ against $5.0$\,bit/s/Hz, so the rule depends on the residuals. Against switching, crosstalk leaves multiplexing only $0.32$ and $0.19$\,bit/s/Hz.

\subsection{Rate Retention}
\begin{figure*}[!t]
\centering
\includegraphics[width=\textwidth]{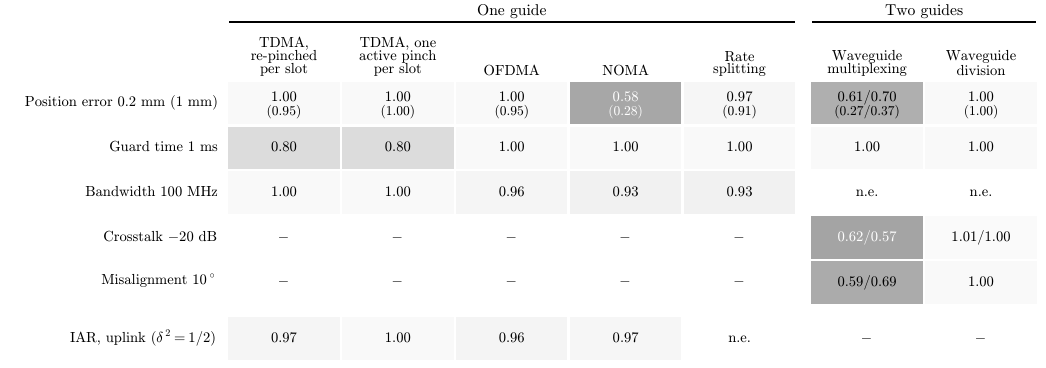}
\caption{Rate retention with one impairment at a time, at $P=20$\,dBm. Two guides: co-located/separated users; $-$: the impairment cannot reach the scheme; n.e.: not evaluated.}
\label{fig:ret}
\end{figure*}
Fig.~\ref{fig:ret} collects the rate retention of seven schemes, the minimum user rate with one impairment divided by that of the same scheme without it. Each impairment is set to one level, and position error also to a second level in parentheses. All schemes in a row see the same level, so the entries of a row can be compared directly. Position error hits the schemes that separate the users through the channel. NOMA keeps $58\,\%$ at $0.2$\,mm. Waveguide multiplexing keeps $61$ to $70\,\%$, because a displaced pinch rotates the phase of its guide as a misaligned RF chain does. TDMA with one active pinch per slot and waveguide division keep their rates. The guard time of $1$\,ms takes $20\,\%$ from the two schemes that reconfigure per slot and nothing from the others. The bandwidth row follows the spread of the pinches rather than the access domain. At $100$\,MHz, per-slot re-pinching and TDMA with one active pinch per slot keep their rates, whereas the schemes with both pinches active lose $4$ to $7\,\%$ with OFDM and $70\,\%$ with the one-tap receiver of Fig.~\ref{fig:bw}. Rate splitting with OFDM, like NOMA, uses one power split for the whole band. Crosstalk and misalignment hit waveguide multiplexing, which keeps $57$ to $69\,\%$, and leave waveguide division almost untouched.

Position error, the guard time, crosstalk, and misalignment thus harm most the schemes that separate their users in the domain that they corrupt, as Table~\ref{tab:matrix} anticipates. The guard time acts on time division, and crosstalk and misalignment act on the spatial nulls. Position error acts on the cancellation of NOMA and on the nulls of waveguide multiplexing. The guard-time and misalignment rows lie on the diagonal of Table~\ref{tab:matrix} by construction, since only per-slot reconfiguration and cross-guide precoding depend on the corrupted quantity. Which impairment hurts a given scheme most depends on the chosen levels, so a column is read only at these levels. On the uplink, each user transmits $20$\,dBm, and NOMA uses time sharing between the two SIC orders. All pinches are identical, with $\delta^2=1/2$, and with per-slot re-pinching both pinches above the served user capture its signal. Compared with the same model without the factor \eqref{eq:iargain}, IAR to first order costs the schemes that keep both pinches active $3$ to $4\,\%$. TDMA with one active pinch per slot loses nothing, as Remark~\ref{rem:iarfree} states.

\section{Conclusion}\label{sec:concl}
This article surveyed PASS multiple access and classified it by its spatial degrees of freedom. One waveguide carries one composite signal per RF chain, so its users can be separated only in time, frequency, power, or guided mode. More waveguides, each with its own feed, add the space domain. On this basis, the article compared the access schemes and identified the impairments that are critical to them. Our unified analytical framework showed that these impairments impact the data rate in three ways: they lower the desired signal strength, waste time resources, or raise the interference power. The first two reduce the rate. The third is the most harmful, since the interference grows with the transmit power, and the rate then hits a ceiling that increasing power cannot overcome. How much a scheme loses therefore depends not only on its channel, but also on how it separates its users.

The two-user benchmark at $20$\,dBm confirmed these dependencies. On one guide, up to eight pinches gathered above the served user add $0.5$\,bit/s/Hz to the minimum user rate per doubling. Per-slot re-pinching leads the best fixed placement by only $2.3\,\%$, and a guard time of $0.11$\,ms per slot of a $10$\,ms frame consumes this lead. Without impairments, NOMA with its best fixed placement is ahead of time division, but a position error of $0.2$\,mm reverses this order. NOMA, which cancels with inaccurate channel information, then keeps $58\,\%$ of its minimum user rate, whereas rate splitting keeps $97\,\%$ by moving the messages into the common stream. With two guides, a crosstalk of $-20$\,dB or a phase error of $10^\circ$ per RF chain takes $31$ to $43\,\%$ from waveguide multiplexing and almost nothing from waveguide division. Over a wide band, the loss depends on the spread of the pinches and on the receiver rather than on the access domain. With one pinch above each user, a one-tap receiver halves its rate at about $40$\,MHz, whereas OFDM loses $4$ to $7\,\%$ at $100$\,MHz.

The analytical and numerical results thus give no universal ranking of the access schemes. The best scheme depends on the operating point, the hardware, and the service target, and it has to be chosen together with the pinch placement, the channel estimation, and the receiver. Open problems remain. The comparison remains to be extended to more users and to the cost of channel estimation. Models of mode crosstalk beyond the leakage at the pinches \cite{Ref_Xu2026MultiModeMDM} and of the uplink beyond first order \cite{Ref_Wang2025Multiport} are still missing. Code-domain access has no PASS instantiation yet, and grant-free access only one \cite{Ref_Shan2026SWANAccess}. Finally, the results await validation by measurement.

\end{document}